\documentclass[journal]{IEEEtran}
\usepackage{cite}
\usepackage{amsmath,amssymb,amsfonts}
\usepackage{multicol}     
\usepackage{graphicx}
\usepackage{textcomp}
\usepackage{svg}
\usepackage{lipsum}    
\usepackage{tabularx}  
\usepackage{booktabs} 
\usepackage{multirow} 
\usepackage{algorithm}  
\usepackage{algorithmicx}  
\usepackage{algpseudocode}

\usepackage{amsmath}  
\usepackage{url}
\usepackage{stackrel}   
\usepackage{pifont}
\usepackage{tikz}
\usepackage{pgfplots}
\usepackage{pgfplotstable} 
\pgfplotsset{compat=1.17} 
\usetikzlibrary{
    shapes.geometric,  
    positioning,       
    fit,               
    arrows.meta,       
    backgrounds,       
    bending,, calc          
}
\usepackage{tabularx} 
\usepackage{amsmath}
\usepackage{booktabs} 
\usepackage{arydshln}       
\usepackage{amssymb}        
\usepackage{multirow}

\usepackage{amsmath, amsfonts, amsthm, stackrel, amssymb}   
 \newtheorem{theorem}{Theorem}[section]

\usepackage{float}
\usepackage{subfigure}

\usepackage{amssymb}

\usepackage{enumerate}

\begin{document}

\title{General Linear-Layer Algorithms for Privacy-Preserving Transformer Inference}

\author{Wenshao Yang,
        Zhenhua Liu
        and Dongdong Yao}

\maketitle

\begin{abstract}
Two-party privacy-preserving Transformer inference today mainly follows one of two paradigms: non-interactive computation over fully homomorphic encryption (FHE), or a hybrid of FHE and secure multi-party computation (MPC). Either way, linear layers account for a large share of the compute and communication budget. Most existing systems attack this cost by designing bespoke plaintext encodings and ciphertext matrix algorithms that minimize the expensive key-switching operations. The price is tight coupling: each scheme only works with its own packing/encoding strategy, and moving data between linear and nonlinear layers still triggers online  ciphertext transmission and ring-field conversion.

We take a different route. Instead of optimizing within a chosen encoding, we design a series of general algorithms for linear layers that can be readily integrated into existing inference systems, based on FHE and additive secret sharing. Our algorithms work with any compatible RLWE-based scheme and plug into both non-interactive FHE and hybrid FHE–MPC pipelines. Output interfaces in proposed algorithms allow them to be composed  with different downstream nonlinear layers. Relative to existing linear layer schemes, our construction removes online key switching, ciphertext transmission, and online ring-field conversion. As a complement, we further propose a fused RNS-CKKS technique that reduces the computational cost of multiplication followed by rotation. We prove correctness and semi-honest security of the proposed algorithms and quantify the resulting savings analytically.
\end{abstract}


\textit{Note: This arXiv version focuses on analytical evaluation. An updated version including end-to-end implementation results and experimental evaluation will be released in a subsequent revision.}

\IEEEpeerreviewmaketitle

\section{Introduction}\label{sec:introduction}

Privacy-preserving large language model (LLM) inference allows a server to evaluate a proprietary Transformer on a client's sensitive input without revealing either the input or the model weights. In two-party systems, linear layers are typically evaluated using fully homomorphic encryption (FHE), while nonlinear functions rely on either FHE or secure multi-party computation (MPC). The resulting overhead arises not only from encrypted arithmetic, but also from key switching, ciphertext transmission, and conversions between cryptographic representations.

Key switching is a major bottleneck in homomorphic matrix multiplication~\cite{park-etal-2025-powerformer,ARION,blb,bolt,Thor,NEURIPS2022_64e2449d}. To reduce this cost, existing systems often co-design plaintext packing with ciphertext matrix-multiplication schedules. BLB and THOR target short sequences, whereas MOAI and ARION emphasize large batches~\cite{blb,Thor,moai,ARION}. NEXUS and Euston similarly optimize preprocessing within their respective inference pipelines~\cite{NEXUS,Euston}. Although effective in their target settings, these optimizations are closely tied to particular packing layouts and execution pipelines, making them difficult to reuse as independent linear-layer components.

This paper addresses this limitation through a family of general linear-layer preprocessing algorithms based on additive masks. The algorithms decompose a server-weight linear layer into input-independent encrypted preprocessing and input-dependent online matrix corrections. The construction is specified through matrix-level relations rather than slot assignments or packing-specific rotation schedules. Attention serves as a running example rather than the scope of the primitive: the same construction applies to query, key, value, and output projections, feed-forward layers, and other server-weight linear transformations. Moreover, the algorithms provide both FHE-output and additive-secret-shared-output interfaces, enabling direct composition with different downstream nonlinear protocols.

Overall, Algorithms~1--7 offer the following advantages:
\begin{itemize}
    \item \textbf{General and reusable preprocessing.} The construction relies only on standard RLWE homomorphism and supports compatible schemes, including CKKS, BFV, and BGV. It is independent of specific plaintext encodings or packing layouts, while FHE ciphertext computing are performed during preprocessing, allowing existing homomorphic matrix-multiplication schemes to be reused~\cite{blb,Thor,moai,ARION}.

    \item \textbf{Lower online cost in non-interactive FHE inference.} The FHE-output algorithms reduce online key switching. The storage-reduced variant further trades $O(m^2)$ plaintext elements of online communication for a substantial reduction in server-side ciphertext storage.

    \item \textbf{Efficient online computation in hybrid FHE--MPC inference.} The covered online computation requires neither FHE ciphertext transmission nor key switching. Instead, the parties exchange masked shares and, when necessary, plaintext-sized corrections.

    \item \textbf{Lower interaction and communication costs for secret-shared outputs.} Compared with Beaver-triple multiplication, which requires two online rounds and $O(m^2)$ communication, the share-output attention algorithms require only one online round and $O(m)$ communication for fixed model width.

    \item \textbf{Direct composition without interactive representation conversion.} Ring-based masked shares enter the linear layer directly. FHE outputs remain in the encrypted domain for downstream nonlinear evaluation, as in BLB~\cite{blb}, while ring-based secret-shared outputs are passed directly to MPC for continued computation over the same ring. Consequently, no interactive ring-to-field or field-to-ring conversion is required at these matching interfaces.
\end{itemize}

\begin{figure*}[htbp]
    \centering
    \includegraphics[width=\textwidth]{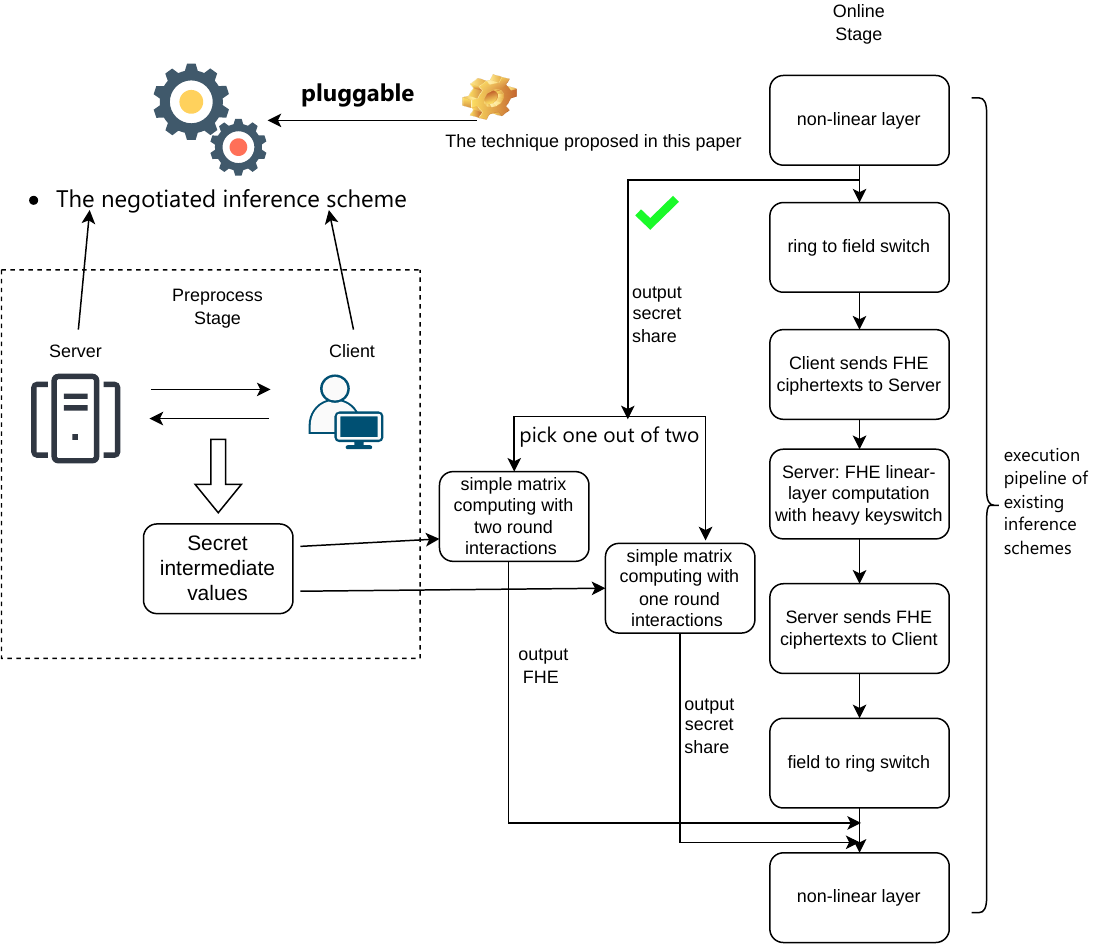}
    \caption{Pluggable offline--online linear-layer preprocessing. A negotiated inference scheme generates masked intermediates offline; online matrix computation then returns either an FHE ciphertext or additive secret shares to the downstream nonlinear pipeline.}
    \label{fig:wide}
\end{figure*}

\subsubsection{Design overview and paper organization}
Figure~\ref{fig:wide} summarizes the central idea: keep the negotiated inference scheme and nonlinear algorithm, but insert a general preprocessing interface between them. Offline preprocessing produces masked intermediate values; online, one- or two-round matrix computation reconstructs the required linear result in the representation expected by the next stage. After related work and preliminaries, Section~\ref{bigframework} derives the general decomposition and its attention instantiation. Section~\ref{sec:general-linear-protocols} develops the FHE-output and secret-shared-output algorithms, explains the residue-domain reconstruction that avoids interactive representation conversion, and analyzes security. Section~\ref{Fused Relinearization and Rotation} introduces the fused relinearization-and-rotation technique. Section~\ref{evaluation} presents the analytical evaluation, and Section~\ref{conclusion} concludes the paper.

\section{Related Work}\label{sec:related-work}

Privacy-preserving machine learning has been built from MPC~\cite{mohassel2017secureml}, FHE~\cite{gilad2016cryptonets}, differential privacy~\cite{abadi2016deep}, and trusted execution environments~\cite{tramer2019slalom}. For cryptographically secure Transformer inference, the main approaches are MPC-based~\cite{li2023mpcformer,2022MPCViT}, FHE-based~\cite{NEXUS,Thor,moai,park-etal-2025-powerformer}, and hybrid FHE--MPC designs~\cite{blb,NEURIPS2022_64e2449d}.

One FHE research direction adapts the model to encrypted computation. PolyTransformer and PowerSoftmax replace nonlinear functions with low-degree approximations to reduce multiplicative depth~\cite{DBLP:conf/icml/ZimermanBDESW24,Zimerman2024PowerSoftmaxTS}, but may require retraining and affect accuracy~\cite{ARION}. Another direction optimizes packing and preprocessing without changing the model. THOR uses diagonal packing for attention~\cite{Thor}; FHEBERTTiny and Tricycle target compact models through encoding and offline preprocessing~\cite{10.1145/3643651.3659893,cryptoeprint:2025/1200}. For higher throughput, NEXUS introduces client-side preprocessing, while MOAI, Arion, and Euston optimize packing, batching, and preprocessing to amortize homomorphic work~\cite{NEXUS,moai,ARION,Euston}. These optimizations remain closely connected to their chosen encoding and operating regimes.

Hybrid systems commonly use FHE for linear layers and MPC for Softmax, normalization, and activations. BOLT~\cite{bolt}, BLB~\cite{blb}, and EncFormer~\cite{EncFormer} represent this design space. BOLT combines optimized ciphertext packing and low-depth matrix-multiplication algorithms with accurate nonlinear approximations~\cite{bolt}. BLB further extends encrypted computation into linear subcomponents of nonlinear functions to reduce conversions and truncation costs~\cite{blb}. EncFormer introduces stage-compatible packing and a complex CKKS--MPC conversion protocol to reduce repacking and conversion communication~\cite{EncFormer}. Nevertheless, linear computation  remains expensive: EncFormer reports that it accounts for more than $75\%$ of execution time in its WAN setting with NVIDIA A100 GPUs~\cite{EncFormer}.

The distinction of this work is therefore not another packing, batching, or matrix-multiplication schedule. NEXUS and Euston optimize preprocessing inside their respective pipelines; we specify a general linear-layer interface whose algebraic decomposition, masked message flow, and FHE-or-share output choices are independent of a concrete packing. Attention demonstrates composition across encrypted matrix products, while the same primitive serves other server-weight linear layers. The fused relinearization-and-rotation technique is orthogonal: it accelerates the key-switching patterns that FHE matrix computation is likely to require, rather than defining the offline–online interface.

\section{Preliminaries}\label{sec:preliminaries}

\subsection{Notations}

Table~\ref{tab:notations} summarizes the notation. Bold lowercase letters (e.g., $\mathbf{x}$) denote vectors and bold uppercase letters (e.g., $\mathbf{X}$) denote matrices. Unless a scheme-specific discussion is stated, $[[\cdot]]$ denotes encryption under the client's public key in any compatible RLWE-based scheme, including CKKS, BFV, and BGV; a matrix may be represented by one or more ciphertexts depending on encoding/packing. The symbols $\langle \cdot \rangle_c$ and $\langle \cdot \rangle_s$ denote additive shares held by the client and server. Matrix operations use the corresponding plaintext domain: exact modular arithmetic for BFV/BGV and approximate fixed-point arithmetic for CKKS.

\begin{table}[t]
\centering
\caption{Cryptographic Notation}
\label{tab:notations}
\small
\begin{tabular}{ll}
\hline
Notation & Meaning \\
\hline

$\mathbb{R},\mathbb{Z}$ &
Sets of real and integer numbers \\

$\mathbb{Z}_{q}$ &
Integer residue space modulo $q$ \\

$N$ &
Polynomial ring degree \\

$q$ &
Ciphertext modulus \\

$R_q$ &
Polynomial ring
$\mathbb{Z}_q[X]/(X^N+1)$ \\

$\Lambda$ &
plaintext polynomial space \\

$[[\mathbf{M}]]$ &
encryption of matrix $\mathbf{M}$ \\

Enc &
encryption \\

$\sigma_{l}^{t},
\sigma_{r}^{t}$ &
Left and right slot rotations by $t$ positions \\

$P,P',P''$ &
Matrix encoding layouts \\

$p(\mathbf{M})$ &
Encoded plaintext polynomial collection of
$\mathbf{M}$ \\

$\ell_P(\mathbf{M})$ &
Number of plaintext polynomials generated by $P$ \\

\hline
\end{tabular}
\end{table}

\subsection{Transformer Model}

The Transformer architecture~\cite{NIPS2017_3f5ee243} is built upon the multi-head self-attention mechanism. Given an input token matrix
\[
\mathbf{X}\in\mathbb{R}^{m\times d},
\]
the query, key, and value matrices are computed as
\[
\mathbf{Q}=\mathbf{X}\mathbf{W}_{q},
\qquad
\mathbf{K}=\mathbf{X}\mathbf{W}_{k},
\qquad
\mathbf{V}=\mathbf{X}\mathbf{W}_{v},
\]
where
\[
\mathbf{W}_{q},
\mathbf{W}_{k},
\mathbf{W}_{v}
\in
\mathbb{R}^{d\times d}
\]
are learnable projection matrices, $m$ denotes the sequence length, and $d$ denotes the hidden dimension.

The matrices $\mathbf{Q}$, $\mathbf{K}$, and $\mathbf{V}$ are partitioned into $H$ attention heads:
\[
\mathbf{Q}^{(h)},
\mathbf{K}^{(h)},
\mathbf{V}^{(h)}
\in
\mathbb{R}^{m\times \frac{d}{H}},
\qquad
h\in\{1,\ldots,H\}.
\]

For each attention head, the attention output is computed as
\[
\mathrm{Att}_{h}(\mathbf{X})
=
\mathrm{Softmax}
\left(
\frac{
\mathbf{Q}^{(h)}
\mathbf{K}^{(h)\top}
}
{\sqrt{d/H}}
\right)
\mathbf{V}^{(h)}.
\]

The outputs of all attention heads are concatenated and projected through the output projection matrix $\mathbf{W}_{o}$:
\[
\mathrm{MHA}(\mathbf{X})
=
\mathrm{Concat}
\left(
\mathrm{Att}_{1}(\mathbf{X}),
\ldots,
\mathrm{Att}_{H}(\mathbf{X})
\right)
\mathbf{W}_{o}.
\]

Residual connections, layer normalization, and a position-wise feed-forward network (FFN) complete a standard Transformer block. Attention is only the running example: the same preprocessing interface applies to feed-forward projections and other server-weight matrix multiplications.

\subsection{Threat Model}

We consider a client holding a private input token matrix $\mathbf{X}$ and a server holding pretrained Transformer weights. The algorithm returns the inference result to the client without revealing the plaintext input to the server or the model parameters to the client; neither party should learn more than its prescribed inputs and outputs.

Following prior work on secure Transformer inference~\cite{blb,ARION,Thor,NEXUS,moai,park-etal-2025-powerformer,bolt,Euston}, we assume static, honest-but-curious (semi-honest) adversaries. Both parties follow the algorithm but may analyze their views, and communication uses secure channels.

\subsubsection{RLWE-Based Homomorphic Encryption}

\paragraph{RLWE Plaintexts and Encryption}
Let \(R=\mathbb{Z}[X]/(X^N+1)\) and \(R_q=R/qR\), where \(q\) denotes the currently used ciphertext modulus. The plaintext representation depends on the underlying scheme. In CKKS, a real or complex message is encoded, using a scaling factor and rounding, into an integer plaintext polynomial \(m\in R\). In BFV and BGV, a plaintext is represented by a polynomial \(m\in R_t\), where \(t\) is the plaintext modulus. SIMD packing may be used to encode multiple message values into one plaintext polynomial, subject to the packing conventions of each scheme.

The RLWE encryption primitive encrypts an encoded plaintext \(m\) under a secret key \(s\) by introducing public randomness and a small error. In its standard compact form, the resulting ciphertext is
\[
\mathsf{ct}=\mathsf{Enc}_s(m)=(c_0,c_1)\in R_q^2.
\]
The decryption phase is computed as \([c_0+c_1s]_q\), from which the plaintext is recovered using the scheme-specific decoding procedure. CKKS recovers an approximation to the original message, whereas BFV and BGV recover the exact plaintext in \(R_t\).

\paragraph{Homomorphic Operations}
RLWE-based ciphertexts support homomorphic addition and multiplication. Ciphertext addition is computationally inexpensive, as it mainly involves coefficient-wise additions in \(R_q\), preserves the standard compact ciphertext form, and introduces only limited noise growth. In contrast, homomorphic multiplication, particularly ciphertext-ciphertext multiplication, incurs substantially higher computational overhead. It requires polynomial multiplications and causes faster noise growth. Moreover, raw ciphertext-ciphertext multiplication generally increases the number of ciphertext components and is therefore usually followed by relinearization to restore the compact form. In CKKS, multiplication also increases the ciphertext scale and is typically followed by rescaling. Consequently, homomorphic multiplication and its associated key-switching and modulus-management operations often dominate the cost of RLWE-based computations.

\subsubsection{Additive Secret Sharing}
We also use 2-out-of-2 additive secret sharing as a standard primitive in hybrid FHE-MPC secure inference algorithms. For an input matrix \(\mathbf{X}\), the client and the server hold shares
\[
\mathbf{X}
=
\left\langle \mathbf{X} \right\rangle_c
+
\left\langle \mathbf{X} \right\rangle_s,
\]
where \(\left\langle \mathbf{X} \right\rangle_c\) is held by the client and \(\left\langle \mathbf{X} \right\rangle_s\) is held by the server. Individually, neither share reveals the plaintext input.

\section{General Linear-Layer Inference and Its Attention Instantiation}\label{bigframework}

Consider an arbitrary linear layer with a client-held input and a server-held weight matrix. NEXUS and Euston demonstrate the efficiency of mask-based preprocessing within their column-packed inference pipelines~\cite{NEXUS,Euston}. Our goal is different: rather than optimize one such pipeline, We construct a series of general linear-layer algorithms whose  logic is independent of specific encoding/packing strategies and RLWE schemes, enabling the optimization of schemes proposed in prior works. The algebraic starting point is
\[
[[\mathbf{X}\mathbf{W}]]
=
[[\mathbf{U}\mathbf{W}]]
+
(\mathbf{X}-\mathbf{U})\mathbf{W},
\]
where $\mathbf{U}$ is a fresh masking matrix generated by the client.

Based on this decomposition, we design Algorithm~\ref{SSFCL} for linear-layer inference in the hybrid FHE--MPC setting. Despite its algebraic simplicity, the algorithm transforms the inference pipeline as shown in Fig.~\ref{fig:wide}: it bypasses several latency-critical operations required by existing pipelines and completes the linear layer with lower communication overhead. The client's decrypted value and the server's locally corrected value form additive shares of the linear output. During online inference, the client sends only a masked input share of plaintext-matrix size, without transmitting FHE ciphertexts or performing key switching. Ciphertext generation and transmission are confined to preprocessing. Algorithm~\ref{SSFCL} returns additive shares over the residue ring $\mathbb{Z}_p$ used by the surrounding MPC protocol, allowing its output to be consumed directly by subsequent MPC-based nonlinear layers. The masked-input reconstruction and the conditions under which the composition requires no interactive ring--field conversion are detailed in Section~V-C.

\begin{algorithm}[htbp]
\caption{ secret-sharing matrix $\mathbf{X}\mathbf{W}$ under hybrid FHE-MPC setting} 
\label{SSFCL}
\begin{algorithmic}[1]
 
\Require
Random masking matrix
$\mathbf{\bar{U}}$
generated by the client,Random masking matrix
$\mathbf{R}$
generated by the server, input secret-shared matrix $\mathbf{X}$ and fully connected layer matrix $\mathbf{W}$.

\Ensure
Server gets $\left\langle \mathbf{X}\mathbf{W} \right\rangle_s$ and Client gets $\left\langle \mathbf{X}\mathbf{W} \right\rangle_c$

\State \textbf{Offline Preprocessing Phase}

\State Client  generates random matrix $\mathbf{\bar{U}}$, then sends the ciphertext $[[\mathbf{\bar{U}}]]$ to server.

\State Server computes $[[\mathbf{\bar{U}}]] \times \mathbf{W}$, then sends  $[[\mathbf{\bar{U}} \mathbf{W}]]-\mathbf{R}$ to client and keeps $\mathbf{R}$ locally.

\State Client decrypts $[[\mathbf{\bar{U}} \mathbf{W}-\mathbf{R}]]$.

\State \textbf{Online Inference Phase}

\State Client sends the mask matrix $\left \langle \mathbf{X}\right \rangle _{c}-\mathbf{\bar{U}}$
to  server.

\State Client obtains  
\[
\left \langle\mathbf{X}\mathbf{W}\right \rangle _{c}=\mathbf{\bar{U}} \mathbf{W}-\mathbf{R}
\]

\State Server obtains 
\[
\left \langle\mathbf{X}\mathbf{W}\right \rangle _{s} = (\left \langle \mathbf{X}\right \rangle _{c}-\mathbf{\bar{U}}+\left \langle \mathbf{X}\right \rangle _{s})\mathbf{W}+\mathbf{R}
\]

\end{algorithmic}
\end{algorithm}

For  attention mechanism  in Transformer model, the same primitive evaluates the $\mathbf{Q}$, $\mathbf{K}$, and $\mathbf{V}$ projections without online rotations. It does not by itself remove the later encrypted attention operations
\[
[[\mathbf{Q}\mathbf{K}^{\top}]]
\quad \text{and} \quad
\mathrm{softmax}\left(
\frac{[[\mathbf{Q}\mathbf{K}^{\top}]]}{\sqrt{d}}
\right)
[[\mathbf{V}]],
\]
which still require ciphertext--ciphertext multiplication and key switching. Algorithm~\ref{alg:secure-attention} therefore composes the general preprocessing principle across the score and value paths.

\subsection{From General Linear Preprocessing to Attention}

We now extend the masked-decomposition idea underlying Algorithm~\ref{SSFCL} to attention, which serves as an instantiation rather than a restriction on the overall design of this work. Its projections use the general linear computing, while its score and value products are further decomposed into precomputable encrypted terms and online masked terms. For a single attention head, define
\[
\mathbf{Q}_{u} = \mathbf{U}\mathbf{W}_{q},
\qquad
\mathbf{K}_{u} = \mathbf{U}\mathbf{W}_{k},
\qquad
\mathbf{V}_{u} = \mathbf{U}\mathbf{W}_{v},
\]
where $\mathbf{W}_{q}$, $\mathbf{W}_{k}$, and $\mathbf{W}_{v}$ denote the query, key, and value projection matrices of a single attention head.

Using the additive decomposition
\[
\mathbf{X} = \mathbf{U} + (\mathbf{X}-\mathbf{U}),
\]
the attention score matrix can be expanded as follows:
\begin{equation}
\label{eq:qk-decomposition}
\begin{aligned}
[[\mathbf{Q}\mathbf{K}^{\top}]]
=
&\;
[[\mathbf{Q}_{u}]]
\times
[[\mathbf{K}_{u}^{\top}]]
\\
&+
[[\mathbf{Q}_{u}]]
\times
\left((\mathbf{X}-\mathbf{U})\mathbf{W}_{k}\right)^{\top}
\\
&+
(\mathbf{X}-\mathbf{U})\mathbf{W}_{q}
\times
[[\mathbf{K}_{u}^{\top}]]
\\
&+
(\mathbf{X}-\mathbf{U})
\mathbf{W}_{q}
\mathbf{W}_{k}^{\top}
(\mathbf{X}-\mathbf{U})^{\top}.
\end{aligned}
\end{equation}

The decomposition exposes the input-independent ciphertext--ciphertext product and the input-dependent mixed terms. Together with precomputed ciphertext transformations, this moves the associated key-switching work from the online phase to preprocessing.

\begin{algorithm}[htbp]
\caption{Two-Party Secure Inference for Attention Mechanism}
\label{alg:secure-attention}
\begin{algorithmic}[1]

\Require
Random masking matrix
$\mathbf{U}\in\mathbb{R}^{m\times {d}}$
generated by the client, input token matrix $\mathbf{X} \in\mathbb{R}^{m\times {d}}$, and server-side single-head attention weights
$\mathbf{W}_{q}^{(h)}$,
$\mathbf{W}_{k}^{(h)}$,
$\mathbf{W}_{v}^{(h)} \in\mathbb{R}^{d\times \frac{d}{H}}$, where $h \in \{1,2,\cdots  ,H\}$.

\Ensure
Encrypted attention output
$[[\mathbf{S}_{5}]]$.

\State \textbf{Offline Preprocessing Phase}

\State The client and server jointly compute the encrypted masked projections:
\[
[[\mathbf{Q}_{u}^{(h)}]]
=
[[\mathbf{U}\mathbf{W}_{q}^{(h)}]]
\quad
[[\mathbf{K}_{u}^{(h)}]]
=
[[\mathbf{U}\mathbf{W}_{k}^{(h)}]]
\]
and 
\[
[[\mathbf{V}_{u}^{(h)}]]
=
[[\mathbf{U}\mathbf{W}_{v}^{(h)}]].
\]
The resulting ciphertexts are held by the server.

\State The server computes:
\[
[[\mathbf{S}_{1}]]
=
[[\mathbf{Q}_{u}^{(h)}]]
\times
[[\mathbf{K}_{u}^{(h) \top}]].
\]

\State The server precomputes transformed ciphertext sets:
\[
C_Q,\quad C_K,\quad C_V.
\]

\State \textbf{Online Inference Phase}

\State The client and server obtain
\[
\mathbf{X}-\mathbf{U}
\]
either via joint computation (as in MPC-based inference) or by the client directly sending it to the server (as in FHE-based inference). 
The resulting value is held by the server.

\State Using ciphertexts in $C_Q$ and $C_K$, the server computes:
\[
[[\mathbf{S}_{2}]]
=
[[\mathbf{Q}_{u}^{(h)}]]
\mathbf{W}_{k}^{(h)\top}
(\mathbf{X}-\mathbf{U})^{\top},
\]
and
\[
[[\mathbf{S}_{3}]]
= 
(\mathbf{X}-\mathbf{U})
\mathbf{W}_{q}^{(h)} [[
\mathbf{K}_{u}^{(h)\top}
]].
\]

\State The server computes:
\[
[[\mathbf{Q}^{(h)} \mathbf{K}^{(h)\top}]]
=
[[\mathbf{S}_{1}]]
+
[[\mathbf{S}_{2}]]
+
[[\mathbf{S}_{3}]]
+
\mathbf{S}_{4}.
\]

\State Securely compute
\[
[[sftmx]]
=
\mathrm{softmax}
\left(
\frac{
[[\mathbf{Q}^{(h)}\mathbf{K}^{(h)\top}]]
}
{\sqrt{d}}
\right).
\]

\State Using ciphertexts in $C_V$, the server computes:
\[
[[\mathbf{S}_{5}]]
=
[[sftmx]] \times \left[
\left[
\mathbf{V}_{u}^{(h)}
+
(\mathbf{X}-\mathbf{U})\mathbf{W}_{v}^{(h)}
\right]
\right].
\]

\State \Return $[[\mathbf{S}_{5}]]$

\end{algorithmic}
\end{algorithm}

\vspace{1em}

This decomposition leads to Algorithm~\ref{alg:secure-attention}, which combines offline preparation of masked projections, their score product, and transformed ciphertext sets with online score evaluation, secure Softmax, and value aggregation. The algorithm retains an FHE output for subsequent homomorphic or BLB-style computation. We detail its procedure below and develop its output-interface variants in the following sections.

\subsection{Correctness and Generality}

We clarify that the proposed attention decomposition in
Eq.~\eqref{eq:qk-decomposition} is both correct and compatible with a broad range of RLWE-based secure attention computation schemes.

\paragraph{Abstract Encoding Model}

To avoid reliance on any particular packing strategy, we model an encoding layout as a mapping from a matrix to a collection of encoded plaintext polynomials:
\[
P:\mathbf{M}\longmapsto p(\mathbf{M}),
\qquad
p(\mathbf{M}) \in \Lambda_{\mathrm{pt}}^{\ell_P(\mathbf{M})},
\]
where $\Lambda_{\mathrm{pt}}$ denotes the plaintext polynomial ring and $\ell_P(\mathbf{M})$ denotes the number of plaintext polynomials produced by the layout $P$.

Let
\[
R=\mathbb{Z}[X]/(X^N+1),
\qquad
R_t=R/tR,
\]
where $t$ is the plaintext modulus used by exact-arithmetic schemes. The plaintext ring is scheme-dependent:
\[
\Lambda_{\mathrm{pt}}=
\begin{cases}
R, & \text{for CKKS},\\[2pt]
R_t, & \text{for BFV and BGV}.
\end{cases}
\]
In CKKS, real or complex messages are scaled, rounded, and encoded as integer polynomials in $R$. In BFV and BGV, plaintexts are represented exactly as polynomials in $R_t$. The value of $\ell_P(\mathbf{M})$ depends on the packing strategy and is not restricted by our algorithms. The ciphertext modulus $q$ remains separate from the plaintext modulus and determines the ring over which ciphertext polynomials are represented.

Different operands and outputs may adopt different encoding layouts:
\[
P':\mathbf{M}\longmapsto p'(\mathbf{M}),
\qquad
P'':\mathbf{M}\longmapsto p''(\mathbf{M}).
\]

The encryption of the encoded plaintext collection \(p(\mathbf{M})\) is denoted by $[[p(\mathbf{M})]]$.

A homomorphic ciphertext-matrix multiplication algorithm is represented as
\[
\varphi :
[[p(\mathbf{A})]]
\times
[[p'(\mathbf{B})]]
\longrightarrow
[[p''(\mathbf{AB})]].
\]

The output encoding layout \(P''\) is assumed to preserve additive homomorphism:
\begin{equation}\label{homomorphic}
p''(\mathbf A+\mathbf B)
=
p''(\mathbf A)
+
p''(\mathbf B),    
\end{equation}
which is satisfied by standard RLWE packing layouts used in secure inference systems.

\paragraph{Correctness}

From
\[
\mathbf{Q}
=
\mathbf{U}\mathbf{W}_q
+
(\mathbf{X}-\mathbf{U})\mathbf{W}_q
=
\mathbf{Q}_u
+
(\mathbf{X}-\mathbf{U})\mathbf{W}_q
\]
and
\[
\mathbf{K}
=
\mathbf{U}\mathbf{W}_k
+
(\mathbf{X}-\mathbf{U})\mathbf{W}_k
=
\mathbf{K}_u
+
(\mathbf{X}-\mathbf{U})\mathbf{W}_k,
\]
we obtain
\[
\mathbf{Q}\mathbf{K}^{\top}
=
\bigl(
\mathbf{Q}_u
+
(\mathbf{X}-\mathbf{U})\mathbf{W}_q
\bigr)
\bigl(
\mathbf{K}_u
+
(\mathbf{X}-\mathbf{U})\mathbf{W}_k
\bigr)^{\top}.
\]
Applying the distributive law of matrix multiplication yields
\begin{equation}
\label{eq:qk-correctness}
\begin{aligned}
\mathbf{Q}\mathbf{K}^{\top}
=
&\;
\mathbf{Q}_u\mathbf{K}_u^{\top}
\\
&+
\mathbf{Q}_u
\bigl(
(\mathbf{X}-\mathbf{U})\mathbf{W}_k
\bigr)^{\top}
\\
&+
(\mathbf{X}-\mathbf{U})\mathbf{W}_q
\mathbf{K}_u^{\top}
\\
&+
(\mathbf{X}-\mathbf{U})
\mathbf{W}_q
\mathbf{W}_k^{\top}
(\mathbf{X}-\mathbf{U})^{\top},
\end{aligned}
\end{equation}
which is exactly Eq.~\eqref{eq:qk-decomposition}. Therefore, the proposed
decomposition is algebraically correct.

\paragraph{Generality}

Since Eq.~\eqref{eq:qk-correctness} is a matrix-level identity, it is independent of the underlying ciphertext representation. Applying the encoding
layout \(P''\) to both sides and using the additive homomorphism of \(P''\) and RLWE, we obtain
\begin{equation} 
\label{eq:generic-he} 
\begin{aligned} [[p''(\mathbf{Q}\mathbf{K}^{\top})]] = &\; \varphi \Big( [[p(\mathbf{Q}_u)]], [[p'(\mathbf{K}_u^{\top})]] \Big) 
\\ &+ \varphi \Big( [[p(\mathbf{Q}_u)]], p'(\mathbf{W}_k^{\top} (\mathbf{X}-\mathbf{U})^{\top}) \Big) 
\\ &+ \varphi \Big( p((\mathbf{X}-\mathbf{U})\mathbf{W}_q), [[p'(\mathbf{K}_u^{\top})]] \Big) 
\\ &+ p''\Big( (\mathbf{X}-\mathbf{U}) \mathbf{W}_q \mathbf{W}_k^{\top} (\mathbf{X}-\mathbf{U})^{\top} \Big). \end{aligned} 
\end{equation}

Therefore, the decomposition is not tied to CKKS or to a particular packing. It can be instantiated with CKKS, BFV, BGV, and other RLWE-based homomorphic encryption schemes that provide the required ciphertext--ciphertext or ciphertext--plaintext operations. CKKS evaluates the identity with bounded approximation error, whereas BFV and BGV evaluate it exactly in their modular plaintext spaces. The only scheme-level requirements are a compatible matrix encoding, the multiplications used by \(\varphi\), and additive homomorphism of the output layout \(P''\). Rotations, relinearization, rescaling, or modulus switching remain internal implementation choices of that routine; our masks, message flow, and output interfaces do not change.

In particular, Algorithm~\ref{alg:secure-attention} can be instantiated within existing CKKS-based secure Transformer inference systems such as NEXUS, EUSTON, Arion, and BLB, and the same construction applies to BFV/BGV implementations with compatible encodings. Since the purpose of introducing \(P\), \(P'\), and \(P''\) is solely to establish packing-independent correctness, we omit these encoding maps in the remainder of this paper for notational simplicity. Accordingly, unless otherwise specified, \([[\mathbf{M}]]\) denotes the ciphertext representation of a matrix \(\mathbf{M}\) using any compatible plaintext encoding layout (including \(P\), \(P'\), or \(P''\)) that satisfies Eq.(~\ref{homomorphic}).

The practical consequence is twofold. First, changing the packing layout may change implementation cost and the number of ciphertexts, but not the mask-based message flow, algebraic decomposition, or output interface. Second, changing among compatible CKKS, BFV, and BGV instantiations changes the plaintext arithmetic and noise management, not the algorithm itself. In this sense, the technique is independent of any particular plaintext encoding, packing method, or RLWE scheme rather than being an optimization of one such choice.

\subsection{Precomputed Ciphertext Transformation Sets}

To minimize online latency, the server precomputes and stores sets of ciphertext variants derived from the original matrices $[[\mathbf{Q}_{u}]]$, $[[\mathbf{K}_{u}]]$, and $[[\mathbf{V}_{u}]]$. We define these sets as follows:
\begin{align*}
C_{Q} & = \{ [[\mathbf{Q}_{u}]]_{1}, [[\mathbf{Q}_{u}]]_{2}, \dots, [[\mathbf{Q}_{u}]]_{n} \}, \\
C_{K} & = \{ [[\mathbf{K}_{u}]]_{1}, [[\mathbf{K}_{u}]]_{2}, \dots, [[\mathbf{K}_{u}]]_{n'} \}, \\
C_{V} & = \{ [[\mathbf{V}_{u}]]_{1}, [[\mathbf{V}_{u}]]_{2}, \dots, [[\mathbf{V}_{u}]]_{n''} \}.
\end{align*}

Each set contains packing-dependent ciphertext variants prepared through rotations, masking, homomorphic additions, or other auxiliary transformations. Precomputing them reduces online key switching but creates substantial storage overhead. The set sizes $n$, $n'$, and $n''$ depend on the packing strategy and system requirements.

\subsection{Instantiation Notes for Secure Attention}

For Algorithm~\ref{alg:secure-attention}, the superscript \((h)\) denotes the \(h\)-th attention head. Define:
\[
\mathbf{S}_{1}
=
\mathbf{Q}_{u}^{(h)}\mathbf{K}_{u}^{(h)\top},
\]
\[
\mathbf{S}_{2}
=
\mathbf{Q}_{u}^{(h)}
\left(
(\mathbf{X}-\mathbf{U})\mathbf{W}_{k}^{(h)}
\right)^{\top},
\]
\[
\mathbf{S}_{3}
=
(\mathbf{X}-\mathbf{U})
\mathbf{W}_{q}^{(h)}
\mathbf{K}_{u}^{(h)\top},
\]
\[
\mathbf{S}_{4}
=
(\mathbf{X}-\mathbf{U})
\mathbf{W}_{q}^{(h)}
\mathbf{W}_{k}^{(h)\top}
(\mathbf{X}-\mathbf{U})^{\top},
\]
and
\[
\mathbf{S}_{5}
=
\mathrm{softmax}
\left(
\frac{\mathbf{Q}^{(h)}\mathbf{K}^{(h)\top}}{\sqrt{d}}
\right)
\mathbf{V}^{(h)}.
\]

Terms that do not depend on the input are computed offline, while only masked input-dependent terms are evaluated online. Line 2 of Algorithm~\ref{alg:secure-attention} may use the preprocessing methods of NEXUS or Euston. It may also use Beaver triples to generate additive shares of the target matrix. The client then encrypts its share, and the server adds its own share locally.

During online evaluation, selected ciphertext rotations are replaced by plaintext rotations. The sets $C_Q$ and $C_K$ store precomputed transformations of $[[\mathbf{Q}_{u}]]$ and $[[\mathbf{K}_{u}^{\top}]]$. The algorithm therefore accelerates the computation.

Although precomputing ciphertext transformations reduces online key switching, ciphertext matrix multiplication in existing inference schemes may require hundreds or even thousands of key-switching operations. Storing all transformed ciphertexts needed to avoid these operations online can therefore make $C_Q$, $C_K$, and $C_V$ impose substantial storage overhead on the server. This motivates the storage-efficient variants developed in the next section.

\section{General Linear-Layer Inference  Algorithms with Flexible Output Interfaces}\label{sec:general-linear-protocols}

Section~\ref{bigframework} introduced the basic preprocessing principle through Algorithms~\ref{SSFCL} and~\ref{alg:secure-attention}, which move input-independent encrypted computation offline. The five algorithms in this section extend these constructions step by step. Algorithm~\ref{tradeoffs2s3FHE} first reduces the ciphertext storage required by Algorithm~\ref{alg:secure-attention}. Algorithm~\ref{tradeoffs2s3} adapts this construction to secret-shared inputs in hybrid FHE--MPC inference, and Algorithm~\ref{tradeoffMPCQK} further replaces the FHE output with additive shares and keeps output-sized corrections local, thereby reducing online communication. The same design process applies to value aggregation: Algorithm~\ref{tradeoffS5} provides an FHE-output construction, from which Algorithm~\ref{tradeoffMPCMHA} derives an additive-share variant with fewer online messages. Across these variants, masked intermediates replace the cached ciphertext sets $C_Q$, $C_K$, and $C_V$, avoiding online key switching in the covered linear computation. Unlike existing designs tied to specific packing/encoding layouts and matrix-multiplication algorithms, all seven algorithms follow the same offline--online principle and expose a common packing- and RLWE-independent interface. Their FHE outputs support downstream homomorphic computation, while their additive-share outputs can be consumed directly by MPC protocols with low online communication.

\subsection{FHE-Output Preprocessing for Non-Interactive FHE Inference}

Algorithm~\ref{tradeoffs2s3FHE} computes the three pre-Softmax score terms without storing $C_Q$ and $C_K$. During preprocessing, the client receives the server-masked intermediates
\[
\mathbf{Q}_{u}^{(h)}\mathbf{W}_{k}^{(h)\top}-\mathbf{R}_{s}^{(h)}
\quad\text{and}\quad
\mathbf{W}_{q}^{(h)}\mathbf{K}_{u}^{(h)\top}-\mathbf{R}_{s'}^{(h)}
\]
The server-held masks protect weight-dependent values from the client. Online, the client sends $\mathbf{X}-\mathbf{U}$ and the correction $\mathbf{B_1}+\mathbf{B_2}$; the server combines them with its encrypted state to reconstruct the encrypted sum of the first three score terms. No online rotation or relinearization is required, and the result remains an FHE ciphertext for subsequent homomorphic computation.

For non-interactive FHE inference, Algorithms~\ref{alg:secure-attention} and~\ref{tradeoffs2s3FHE} target the first Transformer layer. The end-to-end gain therefore depends on the fraction of work covered, and is more visible in shallow models than in deep models with many uncovered layers. Reusing the construction in later layers may require additional interaction and is separate from the base non-interactive claim.

\begin{algorithm}[htbp]
\caption{Storage-Reduced FHE-Output Preprocessing for Computing $[[\mathbf{S}_{1}]]+[[\mathbf{S}_{2}]]+[[\mathbf{S}_{3}]]$ under FHE in Algorithm \ref{alg:secure-attention}}
\label{tradeoffs2s3FHE}
\begin{algorithmic}[1]
 
\Require
$[[\mathbf{Q}_{u}]]$, $[[\mathbf{K}_{u}]]$,$\mathbf{W}_{q}^{(h)}$,
$\mathbf{W}_{k}^{(h)}$ and $\mathbf{U}$ in Algorithm~\ref{alg:secure-attention}. Random masking matrix 
$\mathbf{R}_{s}^{(h)}\in\mathbb{R}^{m\times d}$, $  \mathbf{R}_{s'}^{(h)}\in\mathbb{R}^{d\times m}$, $\mathbf{R}_{c}^{(h)}\in\mathbb{R}^{m\times m}$.
 Input  matrix $\mathbf{X} \in\mathbb{R}^{m\times d}$ and server-side single-head attention weights
$\mathbf{W}_{q}^{(h)}$,
$\mathbf{W}_{k}^{(h)}$,
$\mathbf{W}_{v}^{(h)} \in\mathbb{R}^{d\times \frac{d}{H}}$, where $h \in \{1,2,\cdots ,H\}$.

\Ensure
$[[\mathbf{S}_{1}]]+[[\mathbf{S}_{2}]]+[[\mathbf{S}_{3}]]$.

\State \textbf{Offline Preprocessing Phase}

\State Server computes 
\[
[[\mathbf{Q}_{u}^{(h)}]]\times \mathbf{W}_{k}^{(h) \top } -\mathbf{R}_{s}^{(h)}
\] 
and
\[
\mathbf{W}_{q}^{(h) }\times [[\mathbf{K}_{u}^{(h) \top}]]  -\mathbf{R}_{s'}^{(h)},
\] then sends them to client.

\State Client decrypts ciphertext 
\[
[[\mathbf{Q}_{u}^{(h)}\times \mathbf{W}_{k}^{(h) \top } -\mathbf{R}_{s}^{(h)}]]
\] 
and
\[
[[\mathbf{W}_{q}^{(h) }\times \mathbf{K}_{u}^{(h) \top}  -\mathbf{R}_{s'}^{(h)}]],
\] then encrypts
$\mathbf{R}_{c}^{(h)} $
and sends 
$[[\mathbf{R}_{c}^{(h)}]]$
 to server.

\State Server computes $[[\mathbf{S}_{1}]] \gets [[\mathbf{S}_{1}]] + [[\mathbf{R}_{c}^{(h)}]]$.

\State \textbf{Online Inference Phase}

\State The client computes 

\[ \mathbf{B_1}=
(\mathbf{Q}_{u}^{(h)}\times \mathbf{W}_{k}^{(h) \top } -\mathbf{R}_{s}^{(h)}) \times (\mathbf{X}-\mathbf{U})^{\top}-\mathbf{R}_{c}^{(h)}
\]
and 
\[ \mathbf{B_2}=
(\mathbf{X}-\mathbf{U}) \times (\mathbf{W}_{q}^{(h) }\times \mathbf{K}_{u}^{(h) \top} -\mathbf{R}_{s'}^{(h)}),\]
then sends 
$\mathbf{X}-\mathbf{U}$
and
$\mathbf{B_1}+\mathbf{B_2}$
to server.

\State Server obtains 
\[
\begin{aligned}
[[\mathbf{S}_{1}]]+[[\mathbf{S}_{2}]]+[[\mathbf{S}_{3}]] 
&
=\mathbf{B_1}+\mathbf{B_2} + (\mathbf{X}-\mathbf{U}) \times \mathbf{R}_{s'}^{(h)} \\
& + \mathbf{R}_{s}^{(h)} \times (\mathbf{X}-\mathbf{U})^{\top} + [[\mathbf{S}_{1}]]
\end{aligned}
\]

\State \Return $[[\mathbf{S}_{1}]]+[[\mathbf{S}_{2}]]+[[\mathbf{S}_{3}]]$

\end{algorithmic}
\end{algorithm}

\paragraph{Role of Algorithm~\ref{tradeoffs2s3FHE}}
The algorithm occupies the middle point between caching every ciphertext variant and rotating ciphertexts online. Masked offline coefficients and plaintext-sized online corrections preserve the FHE output and zero online key switching for the covered terms while substantially reducing ciphertext storage on server.

\subsection{FHE- and Share-Output Preprocessing for Hybrid FHE--MPC Inference}

\begin{algorithm}[htbp]
\caption{Storage-Reduced FHE-Output Preprocessing for Computing $[[\mathbf{S}_{1}]]+[[\mathbf{S}_{2}]]+[[\mathbf{S}_{3}]]$ under hybrid FHE-MPC setting} 
\label{tradeoffs2s3}
\begin{algorithmic}[1]
 
\Require
$[[\mathbf{Q}_{u}]]$, $[[\mathbf{K}_{u}]]$,$\mathbf{W}_{q}^{(h)}$,
$\mathbf{W}_{k}^{(h)}$ and $\mathbf{U}$ in Algorithm~\ref{alg:secure-attention}. Random masking matrix $\mathbf{R}_{t}\in\mathbb{R}^{m\times d}$, $\mathbf{R}_{s}^{(h)}\in\mathbb{R}^{m\times d}$, $  \mathbf{R}_{s'}^{(h)}\in\mathbb{R}^{d\times m}$, $\mathbf{R}_{c}^{(h)}\in\mathbb{R}^{m\times m}$. 
 Input  matrix $\mathbf{X} \in\mathbb{R}^{m\times d}$ and server-side single-head attention weights
$\mathbf{W}_{q}^{(h)}$,
$\mathbf{W}_{k}^{(h)}$,
$\mathbf{W}_{v}^{(h)} \in\mathbb{R}^{d\times \frac{d}{H}}$, where $h \in \{1,2,\cdots ,H\}$.

\Ensure
$[[\mathbf{S}_{1}]]+[[\mathbf{S}_{2}]]+[[\mathbf{S}_{3}]]$.

\State \textbf{Offline Preprocessing Phase}

\State Server computes 
\[
[[\mathbf{Q}_{u}^{(h)}]]\times \mathbf{W}_{k}^{(h) \top } -\mathbf{R}_{s}^{(h)}
\] 
and
\[
\mathbf{W}_{q}^{(h) }\times [[\mathbf{K}_{u}^{(h) \top}]]  -\mathbf{R}_{s'}^{(h)},
\] then sends them to client.

\State Client decrypts ciphertext 
\[
[[\mathbf{Q}_{u}^{(h)}\times \mathbf{W}_{k}^{(h) \top } -\mathbf{R}_{s}^{(h)}]]
\] 
and
\[
[[\mathbf{W}_{q}^{(h) }\times \mathbf{K}_{u}^{(h) \top}  -\mathbf{R}_{s'}^{(h)}]],
\] then encrypts 
$\mathbf{R}_{c}^{(h)} $ and sends
$[[\mathbf{R}_{c}^{(h)}]]$  to server.

\State Server computes 
\[
\begin{aligned}
[[D]]
&
=[[\mathbf{Q}_{u}^{(h)}\times \mathbf{W}_{k}^{(h) \top } -\mathbf{R}_{s}^{(h)}]]\times \mathbf{R}_{t}^{\top} + [[\mathbf{R}_{c}^{(h)}]] \\
& + \mathbf{R}_{t} \times [[\mathbf{W}_{q}^{(h) }\times \mathbf{K}_{u}^{(h) \top}  -\mathbf{R}_{s'}^{(h)}]]+[[\mathbf{S}_{1}]]
\end{aligned}
\]

\State \textbf{Online Inference Phase}

\State The client adds the mask matrix $-\mathbf{U}$
to its secret share  $\left \langle \mathbf{X}\right \rangle _{c}$, while the server adds the mask matrix $-\mathbf{R}_{t}$ to its secret share $\left \langle \mathbf{X}\right \rangle _{s}$
. Both parties then simultaneously send their masked shares to each other.

\State Client computes 
\[
\mathbf{E}=(\mathbf{Q}_{u}^{(h)}\times \mathbf{W}_{k}^{(h) \top } -\mathbf{R}_{s}^{(h)}) \times (\mathbf{X}-\mathbf{U}-\mathbf{R}_{t})^{\top}-\mathbf{R}_{c}^{(h)}
\]
and 
\[
\mathbf{F}=(\mathbf{X}-\mathbf{U}-\mathbf{R}_{t}) \times (\mathbf{W}_{q}^{(h) }\times \mathbf{K}_{u}^{(h) \top} -\mathbf{R}_{s'}^{(h)}),
\]
then sends $\mathbf{E}+\mathbf{F}$ to server.

\State Server obtains 
\[
\begin{aligned}
[[\mathbf{S}_{1}]]+[[\mathbf{S}_{2}]]+[[\mathbf{S}_{3}]]
&
= [[D]] +\mathbf{F} + \mathbf{E} + (\mathbf{X}-\mathbf{U}) \times \mathbf{R}_{s'}^{(h)} \\
&   + \mathbf{R}_{s}^{(h)} \times (\mathbf{X}-\mathbf{U})^{\top} 
\end{aligned}
\]


\State \Return $[[\mathbf{S}_{1}]]+[[\mathbf{S}_{2}]]+[[\mathbf{S}_{3}]]$

\end{algorithmic}
\end{algorithm}

The following algorithms cover both attention products and both output representations. Algorithms~\ref{tradeoffs2s3} and~\ref{tradeoffS5} return FHE ciphertexts; Algorithms~\ref{tradeoffMPCQK} and~\ref{tradeoffMPCMHA} return MPC-ready additive shares. Their online linear-layer messages contain no FHE ciphertexts, and share-output variants such as Algorithms~\ref{tradeoffMPCQK} and~\ref{tradeoffMPCMHA} require less plaintext communication, transmitting only $O(m)$ plaintext elements online for fixed Transformer model architecture.

\paragraph{Role of Algorithm~\ref{tradeoffs2s3}}
The offline ciphertext $[[D]]$ collects input-independent and correction terms. Online, the simultaneous exchange of masked shares lets both parties form the same doubly masked input without exposing either original share. The client sends only $\mathbf{E}+\mathbf{F}$, and the server reconstructs the encrypted score terms using $[[D]]$ and its retained masks. In particular, $[[\mathbf{S}_2]]$ and $[[\mathbf{S}_3]]$ are recovered through their sum $[[\mathbf{S}_2]]+[[\mathbf{S}_3]]$, without online rotations or cached transformation sets. Thus the covered computation uses neither online key switching nor online FHE ciphertext transmission, while its FHE output can feed subsequent homomorphic or BLB-style computation.

\begin{algorithm}[htbp]
\caption{Storage- and Communication-Efficient Share-Output Preprocessing for Attention Score Matrices} 
\label{tradeoffMPCQK}
\begin{algorithmic}[1]
 
\Require
$\mathbf{E}$, $\mathbf{F}$, $D$, $\mathbf{U}$, $\mathbf{R}_{t}$, $\mathbf{R}_{s}^{(h)}$, $  \mathbf{R}_{s'}^{(h)}$, $\mathbf{R}_{c}^{(h)}$ and  $\mathbf{X}$ in Algorithm~\ref{tradeoffs2s3}. Random matrix $\mathbf{R}_{ss}^{(h)}\in\mathbb{R}^{m\times m}$.

\Ensure
Server gets $\left\langle \mathbf{Q}\mathbf{K}^{\top} \right\rangle_s$ and Client gets $\left\langle \mathbf{Q}\mathbf{K}^{\top} \right\rangle_c$

\State \textbf{Offline Preprocessing Phase}

\State Server and client jointly execute the step 1 to step 4 in Algorithm~\ref{tradeoffs2s3}.

\State Server sends  $[[D]]-\mathbf{R}_{ss}^{(h)}$ to client and keeps $\mathbf{R}_{ss}^{(h)}$ locally.

\State Client decrypts $[[D-\mathbf{R}_{ss}^{(h)}]]$.

\State \textbf{Online Inference Phase}

\State Client adds the mask matrix $-\mathbf{U}$
to its secret share  $\left \langle \mathbf{X}\right \rangle _{c}$, while server adds the mask matrix $-\mathbf{R}_{t}$ to its secret share $\left \langle \mathbf{X}\right \rangle _{s}$
. Both parties then simultaneously send their masked shares to each other.

\State Client obtains  
\[
\left\langle \mathbf{Q}\mathbf{K}^{\top} \right\rangle_c = \mathbf{E} + \mathbf{F} + D-\mathbf{R}_{ss}^{(h)}
\].

\State Server obtains 
\[
\begin{aligned}
\left\langle \mathbf{Q}\mathbf{K}^{\top} \right\rangle_s
&
= \mathbf{R}_{ss}^{(h)} + (\mathbf{X}-\mathbf{U}) \times \mathbf{R}_{s'}^{(h)} \\
&   + \mathbf{R}_{s}^{(h)} \times (\mathbf{X}-\mathbf{U})^{\top} + \mathbf{S}_{4}
\end{aligned}
\]


\end{algorithmic}
\end{algorithm}

\paragraph{Role of Algorithm~\ref{tradeoffMPCQK}}
Algorithm~\ref{tradeoffMPCQK} is the secret-shared-output counterpart of Algorithm~\ref{tradeoffs2s3}. The server mask $\mathbf{R}_{ss}^{(h)}$ splits $D$ between the parties. Online, the client retains $\mathbf{E}+\mathbf{F}$ in its local share instead of transmitting their sum, while the server combines its mask, correction terms, and $\mathbf{S}_{4}$. The outputs are additive shares of the complete score matrix and can enter MPC Softmax directly. 

\paragraph{Communication and storage implication of Algorithm~\ref{tradeoffMPCQK}}
When attention linear-layer outputs are represented by additive secret shares, as in BOLT~\cite{bolt} and EncFormer~\cite{EncFormer}, directly using Beaver triples for the two attention products requires two online communication rounds and $O(m^2)$ matrix-element communication. Algorithms~\ref{tradeoffMPCQK} and~\ref{tradeoffMPCMHA} instead use one simultaneous masked-share exchange. The covered online communication is $O(m)$ rather than $O(m^2)$. The algorithms also require no stored ciphertext sets $C_Q$, $C_K$, or $C_V$. Their additional client-side storage consists of the prescribed decrypted plaintext intermediates. Because transmitting and receiving units on modern devices can operate independently, the simultaneous bidirectional exchange can be executed as one interaction round.

The value-aggregation path uses the same storage-reduced principle. Here $\mathbf{S}_{5}^{(h)}$ denotes the output of a single head of dimension $m\times \frac{d}{H}$, namely $\mathbf{S}_{5}^{(h)}=\mathrm{Att}_{h}(\mathbf{X})=\mathrm{Softmax}(\cdot)\,\mathbf{V}^{(h)}\in\mathbb{R}^{m\times \frac{d}{H}}$; multi-head composition is addressed below. The Softmax output is represented by shares $\langle \mathbf{X}' \rangle_c$ and $\langle \mathbf{X}' \rangle_s$, and $\mathbf{U}'$ is a fresh preprocessing mask. The ciphertext $[[\mathbf{S}_5]]$ is decomposed as
\begin{equation}
\label{eq:s5-decomp}
\begin{aligned}
[[\mathbf{S}_{5}]]
=
&\;
[[\mathbf{U}']]
[[\mathbf{V}_{u}^{(h)}]]
\\
&+
[[\mathbf{U}']]
(\mathbf{X}-\mathbf{U})\mathbf{W}_{v}^{(h)}
\\
&+
(\mathbf{X}'-\mathbf{U}')
[[\mathbf{V}_{u}^{(h)}]]
\\
&+
(\mathbf{X}'-\mathbf{U}')
(\mathbf{X}-\mathbf{U})\mathbf{W}_{v}^{(h)}.
\end{aligned}
\end{equation}

Define the four resulting terms as
\[
\hat{S}_{1} = \mathbf{U}'
\mathbf{V}_{u}^{(h)}
\]
\[
\hat{S}_{2}
=
\mathbf{U}'
(\mathbf{X}-\mathbf{U})\mathbf{W}_{v}^{(h)}
\]
\[
\hat{S}_{3}
=
(\mathbf{X}'-\mathbf{U}')
\mathbf{V}_{u}^{(h)}
\]
\[
\hat{S}_{4}=(\mathbf{X}'-\mathbf{U}')
(\mathbf{X}-\mathbf{U})\mathbf{W}_{v}^{(h)}.
\]

The mixed terms $\hat{S}_{2}$ and $\hat{S}_{3}$ originally require $C_V$. In Algorithm~\ref{tradeoffS5}, the underlying value $G$ gathers the offline value-path and correction terms and is retained as $[[G]]$; the client receives only a server-masked value projection. Online, the masked Softmax shares and plaintext corrections $\mathbf{J}$ and $\mathbf{P}$ reconstruct the encrypted mixed terms without key switching. 

\noindent\textbf{Remark (Extension to the multi-head output projection).}
The conventional multi-head output concatenates all heads and applies $\mathbf{W}_{o}\in\mathbb{R}^{d\times d}$:
\[
\mathrm{MHA}(\mathbf{X})
=
\mathrm{Concat}\big(\mathbf{S}_{5}^{(1)},\cdots,\mathbf{S}_{5}^{(H)}\big)\,\mathbf{W}_{o}.
\]
Partitioning $\mathbf{W}_{o}$ row-wise into $H$ blocks $\mathbf{W}_{o}^{(h)}\in\mathbb{R}^{\frac{d}{H}\times d}$, $h\in\{1,\cdots,H\}$, gives
\[
\mathrm{MHA}(\mathbf{X})
=
\sum_{h=1}^{H}\mathbf{S}_{5}^{(h)}\,\mathbf{W}_{o}^{(h)}.
\]
Each block $\mathbf{W}_{o}^{(h)}$ is held in plaintext by the server. By linearity of Eq.~\eqref{eq:s5-decomp}, the projected head output $\mathbf{S}_{5}^{(h)}\mathbf{W}_{o}^{(h)}$ is obtained by replacing each $\hat{S}_{i}$ ($i\in\{1,2,3,4\}$) with $\hat{S}_{i}\mathbf{W}_{o}^{(h)}$:
\[
\begin{aligned}
\hat{S}_{1}\mathbf{W}_{o}^{(h)}
&
=
\mathbf{U}'\big(\mathbf{V}_{u}^{(h)}\mathbf{W}_{o}^{(h)}\big),
\\
\hat{S}_{2}\mathbf{W}_{o}^{(h)}
&
=
\mathbf{U}'(\mathbf{X}-\mathbf{U})\big(\mathbf{W}_{v}^{(h)}\mathbf{W}_{o}^{(h)}\big),
\\
\hat{S}_{3}\mathbf{W}_{o}^{(h)}
&
=
(\mathbf{X}'-\mathbf{U}')\big(\mathbf{V}_{u}^{(h)}\mathbf{W}_{o}^{(h)}\big),
\\
\hat{S}_{4}\mathbf{W}_{o}^{(h)}
&
=
(\mathbf{X}'-\mathbf{U}')(\mathbf{X}-\mathbf{U})\big(\mathbf{W}_{v}^{(h)}\mathbf{W}_{o}^{(h)}\big).
\end{aligned}
\]
The server uses the fused ciphertext $[[\mathbf{V}_{u}^{(h)}\mathbf{W}_{o}^{(h)}]]$, computed offline as $[[\mathbf{V}_{u}^{(h)}\mathbf{W}_{o}^{(h)}]]=[[\mathbf{V}_{u}^{(h)}]]\times\mathbf{W}_{o}^{(h)}$, and the fused plaintext weight $\mathbf{W}_{v}^{(h)}\mathbf{W}_{o}^{(h)}$. The fused ciphertext covers $\hat{S}_{1}\mathbf{W}_{o}^{(h)}$ and $\hat{S}_{3}\mathbf{W}_{o}^{(h)}$, while the fused plaintext weight covers $\hat{S}_{2}\mathbf{W}_{o}^{(h)}$ and $\hat{S}_{4}\mathbf{W}_{o}^{(h)}$. Algorithm~\ref{tradeoffS5} then runs unchanged; its projected mixed-term component is $[[(\hat{S}_{2}+\hat{S}_{3})\mathbf{W}_{o}^{(h)}]]$. Summing the four terms and all heads gives
\[
\begin{aligned}
\mathrm{MHA}(\mathbf{X})= 
\sum_{h=1}^{H}\bigg(\sum_{i=1}^{4}\hat{S}_{i}\bigg)\mathbf{W}_{o}^{(h)}&
=\sum_{h=1}^{H}\mathbf{S}_{5}^{(h)}\times\mathbf{W}_{o}^{(h)}
\\ = \sum_{h=1}^{H}\mathrm{MHA}(\mathbf{X})^{(h)}
\end{aligned}
\]

This summation yields the complete encrypted output $[[\mathrm{MHA}(\mathbf{X})]]$. Thus concatenation and output projection are absorbed into the value path without an additional interaction round or an increase in key-switching operations; only the plaintext weight dimensions grow from $\mathbb{R}^{d\times \frac{d}{H}}$ to $\mathbb{R}^{d\times d}$. Together, the FHE- and share-output variants eliminate $C_Q$, $C_K$, and $C_V$ in favor of the explicitly accounted corrections $\mathbf{E}+\mathbf{F}$, $\mathbf{J}$, and $\mathbf{P}$.

\begin{algorithm}[htbp]
\caption{Storage-Reduced FHE-Output Preprocessing for Computing $[[\hat{S}_{1}]]+[[\hat{S}_{2}]]+[[\hat{S}_{3}]]$ under hybrid FHE-MPC setting} 
\label{tradeoffS5}
\begin{algorithmic}[1]
 
\Require
$[[\mathbf{V}_{u}^{(h)}]]$, $(\mathbf{X}-\mathbf{U}-\mathbf{R}_{t})$,
$\mathbf{W}_{v}^{(h)}$ and $\mathbf{R}_t$ in  Algorithm~\ref{tradeoffs2s3} and Algorithm~\ref{alg:secure-attention}. Random masking matrix $\hat{\mathbf{R}}_{c}\in\mathbb{R}^{m\times d}$, $\mathbf{U}'\in\mathbb{R}^{m\times m}$, $  \hat{\mathbf{R}}_{s}^{(h)}\in\mathbb{R}^{m\times d}$, $\hat{\mathbf{R}}_{t}\in\mathbb{R}^{m\times m}$, $\hat{\mathbf{R}}^{ (h)}_{c'}\in\mathbb{R}^{m\times \frac{d}{H}}$.
Attention mechanism input  matrix $\mathbf{X} \in\mathbb{R}^{m\times d}$ and the output $\mathbf{X'} \in\mathbb{R}^{m\times m}$ of Softmax function.

\Ensure
$[[\hat{S}_{1}]]+[[\hat{S}_{2}]]+[[\hat{S}_{3}]]$.

\State \textbf{Offline Preprocessing Phase}

\State Client encrypt $\mathbf{U}'$,  $\hat{\mathbf{R}}_{c}$ and $\hat{\mathbf{R}}^{ (h)}_{c'}$, then send $[[\mathbf{U}']]$,  $[[\hat{\mathbf{R}}_{c}]]$ and $[[\hat{\mathbf{R}}^{ (h)}_{c'}]]$ to server.

\State Server computes 

\[
\begin{aligned}
[[G]]
&
= [[\hat{S}_{1}]]+[[\mathbf{U}']]\times (\mathbf{R}_{t}\mathbf{W}_{v}^{(h)}) + [[\hat{\mathbf{R}}_{c}]] \times \mathbf{W}_{v}^{(h)} \\
&   + \hat{\mathbf{R}}_{t} \times [[\mathbf{V}_{u}^{(h)}]] + [[\hat{\mathbf{R}}_{c'}^{(h)}]]
\end{aligned}
\]

\State Server computes and sends $[[\mathbf{V}^{(h)}_{u}-\hat{\mathbf{R}}_{s}^{(h)}]]$
to client, client decrypt it after reception.

\State \textbf{Online Inference Phase}

\State The client adds the mask matrix $-\mathbf{U'}$
to its secret share  $\left \langle \mathbf{X'}\right \rangle _{c}$, while the server adds the mask matrix $-\hat{\mathbf{R}}_{t}$ to its secret share $\left \langle \mathbf{X'}\right \rangle _{s}$
. Both parties then simultaneously send their masked shares to each other.

\State Client computes and  sends
\[
\mathbf{J}=\mathbf{U'} \times (\mathbf{X}-\mathbf{U}-\mathbf{R}_{t})-\hat{\mathbf{R}}_{c}
\]
and 
\[
\mathbf{P}=(\mathbf{X'}-\mathbf{U'}-\hat{\mathbf{R}}_{t}) \times ( \mathbf{V}_{u}^{(h)} -\hat{\mathbf{R}}_{s}^{(h)}) -\hat{\mathbf{R}}_{c'}^{(h)}
\]
to server.

\State Server obtains 

\[
\begin{aligned}
[[\hat{S}_{1}]]+[[\hat{S}_{2}]]+[[\hat{S}_{3}]]
&
= [[G]] + \mathbf{J} \times \mathbf{W}_{v}^{(h)} \\
&   + \mathbf{P} + (\mathbf{X'}-\mathbf{U'}-\hat{\mathbf{R}}_{t}) \times \hat{\mathbf{R}}_{s}^{(h)}
\end{aligned}
\]

\State \Return $[[\hat{S}_{1}]]+[[\hat{S}_{2}]]+[[\hat{S}_{3}]]$

\end{algorithmic}
\end{algorithm}

\paragraph{Online role of Algorithm~\ref{tradeoffS5}}
The correction $\mathbf{J}$ carries the masked value-projection contribution, and $\mathbf{P}$ carries the masked Softmax-dependent contribution. Together with $[[G]]$ and the retained masks, they let the server reconstruct the FHE result without online key switching or ciphertext transmission. The output can therefore continue directly into homomorphic or BLB-style FHE-extended computation.

\begin{algorithm}[htbp]
\caption{Storage- and Communication-Efficient secret-sharing matrix $\mathrm{MHA}(\mathbf{X})$ computing under hybrid FHE-MPC setting} 
\label{tradeoffMPCMHA}
\begin{algorithmic}[1]
 
\Require
$[[\mathbf{V}_{u}^{(h)}]]$, $(\mathbf{X}-\mathbf{U}-\mathbf{R}_{t})$,
$\mathbf{W}_{v}^{(h)}$, $\mathbf{R}_t$, $\hat{\mathbf{R}}_{c}$, $\mathbf{U}'$, $  \hat{\mathbf{R}}_{s}^{(h)}$, $\hat{\mathbf{R}}_{t}$, $\hat{\mathbf{R}}^{ (h)}_{c'}$, $\mathbf{X} $ and $\mathbf{X'} $ in Algorithm~\ref{tradeoffS5}. Random matrix $\hat{\mathbf{R}}_{ss}^{(h)}\in\mathbb{R}^{m\times \frac{d}{H}}$.

\Ensure
Server gets $\left\langle \mathrm{MHA}(\mathbf{X})^{(h)} \right\rangle_s$ and Client gets $\left\langle \mathrm{MHA}(\mathbf{X})^{(h)} \right\rangle_c$.

\State \textbf{Offline Preprocessing Phase}

\State Client encrypt $\mathbf{U}'$ and $\hat{\mathbf{R}}_{c}$, then send $[[\mathbf{U}']]$and $[[\hat{\mathbf{R}}_{c}]]$ to server.

\State Server computes 
\[
\begin{aligned}
[[G]]
&
= [[\hat{S}_{1}\mathbf{W}_{o}^{(h)}]]+[[\mathbf{U}']]\times (\mathbf{R}_{t}\mathbf{W}_{v}^{(h)}\mathbf{W}_{o}^{(h)})  \\
&   + \hat{\mathbf{R}}_{t} \times [[\mathbf{V}_{u}^{(h)}\mathbf{W}_{o}^{(h)}]] + [[\hat{\mathbf{R}}_{c}]] \times \mathbf{W}_{v}^{(h)}\mathbf{W}_{o}^{(h)}
\end{aligned}
\]

\State Server sends  $[[G]]-\hat{\mathbf{R}}_{ss}^{(h)}$ and $[[\mathbf{V}^{(h)}_{u}\mathbf{W}_{o}^{(h)}-\hat{\mathbf{R}}_{s}^{(h)}]]$ to client, then keeps $\hat{\mathbf{R}}_{ss}^{(h)}$ locally.

\State Client decrypts $[[G-\hat{\mathbf{R}}_{ss}^{(h)}]]$ and $[[\mathbf{V}^{(h)}_{u}\mathbf{W}_{o}^{(h)}-\hat{\mathbf{R}}_{s}^{(h)}]]$.

\State \textbf{Online Inference Phase}

\State The client adds the mask matrix $-\mathbf{U'}$
to its secret share  $\left \langle \mathbf{X'}\right \rangle _{c}$, while the server adds the mask matrix $-\hat{\mathbf{R}}_{t}$ to its secret share $\left \langle \mathbf{X'}\right \rangle _{s}$
. Both parties then simultaneously send their masked shares to each other.

\State Client computes and  sends
\[
\mathbf{J}=\mathbf{U'} \times (\mathbf{X}-\mathbf{U}-\mathbf{R}_{t})-\hat{\mathbf{R}}_{c}
\] to server. 

\State Client obtains  
\[
\begin{aligned}
\left\langle \mathrm{MHA}(\mathbf{X})^{(h)} \right\rangle_c &= (\mathbf{X'}-\mathbf{U'}-\hat{\mathbf{R}}_{t}) \times ( \mathbf{V}_{u}^{(h)}\mathbf{W}_{o}^{(h)}-\hat{\mathbf{R}}_{s}^{(h)}) \\ 
& -\hat{\mathbf{R}}_{c'}^{(h)}+G-\hat{\mathbf{R}}_{ss}^{(h)}
\end{aligned}
\]
\State Server obtains 
\[
\begin{aligned}
\left\langle \mathrm{MHA}(\mathbf{X})^{(h)} \right\rangle_s
&
= \hat{\mathbf{R}}_{ss}^{(h)} + \mathbf{J} \times \mathbf{W}_{v}^{(h)}\mathbf{W}_{o}^{(h)}+\hat{S}_{4}\mathbf{W}_{o}^{(h)} \\
&   + (\mathbf{X'}-\mathbf{U'}-\hat{\mathbf{R}}_{t}) \times \hat{\mathbf{R}}_{s}^{(h)}
\end{aligned}
\]


\end{algorithmic}
\end{algorithm}

\paragraph{Role of Algorithm~\ref{tradeoffMPCMHA}}
Algorithm~\ref{tradeoffMPCMHA} returns additive shares of each projected head output. The mask $\hat{\mathbf{R}}_{ss}^{(h)}$ protects the offline server value and provides output-share randomness. Online, only $\mathbf{J}$ is sent; the Softmax-dependent correction remains in the client share. Compared with Algorithm~\ref{tradeoffS5}, which sends both $\mathbf{J}$ and $\mathbf{P}$, this further reduces online plaintext communication. The shares enter layer normalization, activation, or another MPC nonlinear algorithm directly, avoiding both representation conversions at the covered interface. Algorithms~\ref{tradeoffs2s3} and~\ref{tradeoffS5} remain necessary when downstream computation should stay in the FHE domain.

\subsection{Why the First Seven Algorithms Need No Interactive Ring--Field Conversion}\label{sec:no-ring-field-conversion}

Interactive ring-to-field and field-to-ring algorithms are commonly introduced when MPC shares over a residue ring must be re-encoded for FHE linear computation, or when an FHE result must be converted back into MPC shares. In the established pipeline used by default in the existing literature, when one party needs to combine the shares of $\mathbf{X}$ locally, one additive share of $\mathbf{X}$ is always encrypted under FHE, so the input shares cannot be directly combined into a complete masked plaintext input. In contrast, the first seven algorithms all make the complete input available in masked form, such as $\mathbf{X}-\mathbf{U}$ or $\mathbf{X}-\mathbf{U}-\mathbf{R}_t$, during execution. Provided that any potential wrap-around in the original residue ring can be resolved during the reconstruction of these masked values, they can be encoded directly into the selected RLWE plaintext domain for linear computation, without an interactive ring-to-field conversion.

Let the nonlinear layer use additive sharing over $\mathbb{Z}_p$, where
$\mathbb{Z}_p$ denotes the residue-class ring of integers modulo $p$, and
$p$ need not be prime. Consider the exchange in
Algorithms~\ref{tradeoffs2s3} and~\ref{tradeoffMPCQK}. The matrices
$\mathbf{X}$, $\mathbf{U}$, and $\mathbf{R}_t$ used in these algorithms are
integer matrices whose entries satisfy
\[
-\frac{p}{4}\leq\mathbf{X}<\frac{p}{4},
\qquad
0\leq\mathbf{U},\mathbf{R}_t<\frac{p}{8},
\]
where all matrix inequalities are understood entrywise. Consequently, the
doubly masked matrix
\[
\mathbf{T}=\mathbf{X}-\mathbf{U}-\mathbf{R}_t
\]
satisfies
\[
-\frac{p}{2}<\mathbf{T}<\frac{p}{4}
\]
entrywise. The client and the server each add a mask only to their own
share: the client computes
$\left\langle\mathbf{X}\right\rangle_c-\mathbf{U}$, while the server
computes $\left\langle\mathbf{X}\right\rangle_v-\mathbf{R}_t$. Without
performing any additional computation, the client and the server directly
send their masked shares to each other. When the algorithm subsequently
combines the client's and server's shares to compute $\mathbf{T}$ or
$\mathbf{X}-\mathbf{U}$, the corresponding shares are added entrywise as
ordinary integers. The entries of the resulting matrix are then corrected
as follows, where $s$ denotes an arbitrary entry of that matrix:
\[
s\leftarrow
\begin{cases}
s-p,   & \dfrac{p}{2}<s<\dfrac{5p}{4},\\[4pt]
s-2p,  & \dfrac{3p}{2}<s<2p,\\[4pt]
s,     & \text{otherwise}.
\end{cases}
\]
No further modular reduction is performed until the end of the algorithm's
execution. The bounds on $\mathbf{X}$ and the masks ensure that any hidden
multiple of $p$ can be removed unambiguously. The same procedure applies to
the Softmax-share exchange using $\mathbf{U}'$ and $\hat{\mathbf{R}}_t$ in
Algorithms~\ref{tradeoffS5} and~\ref{tradeoffMPCMHA}.

This conditional subtraction is local arithmetic in $\mathbb{Z}_p$, not a representation conversion. The reconstructed masked value can be encoded directly into the selected RLWE plaintext domain, and all subsequent matrix identities remain unchanged. Algorithms~\ref{tradeoffs2s3FHE},~\ref{tradeoffs2s3}, and~\ref{tradeoffS5} return ciphertexts to computation that remains homomorphic; Algorithms~\ref{SSFCL},~\ref{tradeoffMPCQK}, and~\ref{tradeoffMPCMHA} return shares directly to MPC. Algorithm~\ref{alg:secure-attention} likewise accepts the masked input produced either jointly or directly. Consequently, the matching composition requires neither an interactive ring-to-field switch at the input nor a field-to-ring switch at the output.

\subsection{Security Analysis}\label{sec:security-analysis}

We prove security in the real/ideal simulation paradigm for Algorithm~\ref{SSFCL}, the output-interface algorithms in Algorithms~\ref{tradeoffs2s3FHE}--\ref{tradeoffMPCMHA}, and the linear fragments of Algorithm~\ref{alg:secure-attention}. Nonlinear operations are treated as separate secure subalgorithms. The proof shows that the masked messages introduced by the storage--communication trade-off reveal neither the client input nor the server weights beyond the prescribed output interface.

\subsubsection{Security Model and Cryptographic Assumptions}

We consider a static, semi-honest, probabilistic polynomial-time adversary corrupting either the client or the server. The two parties do not collude, follow every algorithm step, and communicate over authenticated secure channels. The client generates the key pair for the selected RLWE scheme; the server receives only the public and evaluation keys and never receives the secret key. Matrix dimensions, model architecture, number of heads, scheme parameters, communication sizes, and the offline/online timing of messages are public algorithm metadata. No claim is made to hide this metadata.

The selected CKKS, BFV, or BGV instantiation is assumed to be IND-CPA secure with the evaluation keys made available by the algorithm. Additive masks are sampled uniformly from the corresponding finite plaintext domain after encoding. For CKKS, we express the simulation argument over discretized fixed-point representatives; bounded approximation error is a correctness and precision issue and is not used as a privacy mechanism. BFV and BGV instead provide exact arithmetic over their modular plaintext domains.

Each mask must be held by the party whose private value it protects and must be hidden from the other party. In particular, the client generates the input masks $\mathbf{U}$ and $\mathbf{U}'$ and the client-side correction masks, while the server generates the masks protecting weight-dependent intermediates, including $\mathbf{R}_{s}^{(h)}$, $\mathbf{R}_{s'}^{(h)}$, $\mathbf{R}_{t}$, $\hat{\mathbf{R}}_{s}^{(h)}$, $\hat{\mathbf{R}}_{t}$, and the output-splitting masks. A mask is fresh for each independent algorithm execution, layer, and head unless the algorithm explicitly uses the same mask in several algebraically correlated terms. Such correlated terms are simulated jointly from one mask, never as independently random messages. Reusing a mask in unrelated revealed messages is outside the security claim.

For a matrix $A$ and an independently uniform mask $M$ over a finite additive group $\mathcal{G}$, the one-time-mask property gives
\[
\Pr[A-M=z\mid A=a,V=v]=|\mathcal{G}|^{-1}
\qquad\text{for every }z\in\mathcal{G},
\]
whenever $M$ is independent of the already available view $V$. Thus, $A-M$ is uniform and independent of $A$ and $V$. This is a joint-distribution argument: all messages protected by the same mask are derived after sampling that mask once, which preserves the correlations present in the real algorithm.

\subsubsection{Ideal Functionalities and Output Interfaces}

Let $Y$ denote the value prescribed by the selected algorithm, such as a general linear-layer output, the encrypted score terms, the complete score matrix, or a projected attention-head output. An ideal trusted party receives the client input, or the two input shares in the hybrid setting, and the server weights, computes $Y$, and provides only the interface specified below.

\paragraph{FHE-output functionality}
The functionality stores $Y$ and returns an opaque computation handle to the server; neither party receives a plaintext output. In the real algorithm, this handle is instantiated by the corresponding RLWE ciphertext retained by the server. Later homomorphic operations are represented by queries on stored handles. Modeling the ciphertext as an internal encrypted state, rather than as a plaintext value disclosed to the server, matches its actual use in the FHE and BLB-style pipelines.

\paragraph{Secret-shared-output functionality}
The functionality samples one additive share uniformly and sets the other share so that the two shares sum to $Y$. It returns one share to each party. These shares are outputs because they are consumed directly by the following MPC subalgorithm; individually, neither share reveals $Y$.

Softmax and other nonlinear operations are not re-proven here. They are modeled by their own ideal functionalities, and the security of the complete hybrid construction follows by modular composition.

\subsubsection{Simulator for a Corrupted Server}

The server simulator $\mathsf{Sim}_S$ receives the server input, the public keys and parameters, and the prescribed ideal handle or server output share. It constructs the server view as follows.

\begin{enumerate}
    \item \textbf{Client ciphertexts.} For every ciphertext sent by the client, including encryptions of input masks and client-side correction masks, the simulator produces a fresh encryption of the zero plaintext with the same dimensions and modulus level. A sequence of IND-CPA hybrids replaces the real client ciphertexts by these simulated ciphertexts.
    \item \textbf{Masked input shares.} Every client-to-server masked input share is sampled uniformly from its matrix domain. In the real view, $\mathbf{U}$ or $\mathbf{U}'$ is uniform, unknown to the server, and independent of the client input share, so the one-time-mask property gives exactly the same distribution. The simulator derives correlated occurrences from the same simulated mask.
    \item \textbf{Plaintext corrections.} The messages $\mathbf{B_1}+\mathbf{B_2}$ and $\mathbf{E}+\mathbf{F}$ are simulated using the client-side mask $\mathbf{R}_{c}^{(h)}$, which the server knows only in encrypted form. Likewise, $\mathbf{J}$ and $\mathbf{P}$ are simulated using their independent client-side correction masks. After the ciphertext hybrids, these masked messages are uniform in the real view and can therefore be replaced by the simulated values without changing the distribution. All correlated local values are then computed exactly as in the algorithm.
    \item \textbf{Output consistency.} For an FHE-output algorithm, the simulated encrypted state is used only as the real-world handle and no plaintext output is produced. For a secret-shared-output algorithm, the simulator uses the ideal server share to choose the final output-splitting mask, so that the local output equals the prescribed share. Since this share is itself uniform, the programmed mask remains uniformly distributed.
\end{enumerate}

Consequently, the real and simulated server views are computationally indistinguishable. The only computational replacements are RLWE ciphertexts; the masked-share and plaintext-correction replacements are exact in the discretized plaintext domain.

\subsubsection{Simulator for a Corrupted Client}

The client simulator $\mathsf{Sim}_C$ receives the client input or input share, the client key pair, and the prescribed ideal client output share when applicable. It cannot rely on IND-CPA security for ciphertexts that the client itself decrypts. Instead, it simulates their decrypted plaintexts directly using the server-held masks.

\begin{enumerate}
    \item \textbf{Masked weight-dependent intermediates.} Every server-to-client value of the form of a weight-dependent intermediate minus $\mathbf{R}_{s}^{(h)}$ or $\mathbf{R}_{s'}^{(h)}$ is sampled uniformly. The same procedure applies to the masked cached value projection and to values masked by $\mathbf{R}_{ss}^{(h)}$ or $\hat{\mathbf{R}}_{ss}^{(h)}$. Because the corresponding server mask is uniform and unknown to the client, the real decrypted value is independent of the server weights and preprocessed intermediates.
    \item \textbf{Masked server shares.} The shares received after the server adds $-\mathbf{R}_{t}$ or $-\hat{\mathbf{R}}_{t}$ are sampled uniformly. Although the client can combine such a share with its own masked share to obtain the doubly masked value used by the algorithm, it does not know the server mask and therefore cannot remove that mask to recover the original share or input.
    \item \textbf{Decryption consistency.} For every simulated server ciphertext, the simulator encrypts the simulated masked plaintext under the client public key with correctly distributed randomness. Decryption by the client therefore yields exactly the simulated plaintext. Outgoing client messages are computed from the client input and the simulated masks as in the real algorithm.
    \item \textbf{Output consistency.} In a secret-shared-output algorithm, the simulator chooses the final masked server-to-client value so that the client obtains its ideal output share. One additive degree of freedom in the output-splitting mask always permits this choice, and uniformity of the ideal client share implies that the simulated message has the real uniform distribution.
\end{enumerate}

The client simulation therefore reproduces both the messages and their joint correlations without access to the server weights. Its distribution is statistically identical in the exact-group model, up to the explicitly excluded approximation error when CKKS is used.

\subsubsection{Security Theorem and Hybrid Argument}

\begin{theorem}
Under the IND-CPA security of the deployed RLWE instantiation, the one-time-mask assumptions above, and secure ideal functionalities for the nonlinear subalgorithms, the preprocessing algorithms securely realize their corresponding FHE-output or secret-shared-output functionalities against a static semi-honest adversary corrupting either party.
\end{theorem}

\begin{proof}
For a corrupted server, begin with the real execution and replace client-generated ciphertexts one at a time by encryptions of zero. Each replacement is computationally indistinguishable by IND-CPA security, even with the public evaluation keys. Then replace the masked client shares and plaintext corrections by the simulator's values. The one-time-mask property makes this second step distribution-preserving when all values derived from a common mask are replaced jointly. For a corrupted client, first replace every decrypted server message protected by a server-held mask with a uniform value, and then form ciphertexts that decrypt to those values. The mask property again makes the replacement distribution-preserving. Finally, in a secret-shared-output execution, program the remaining output-splitting mask to match the ideal share; uniformity of that share makes this programming statistically indistinguishable from the real mask distribution.

Thus, for each corrupted party $P\in\{C,S\}$, the simulator produces a view satisfying
\[
(\mathsf{view}_{P}^{\mathrm{real}},\mathsf{out}_{\bar P}^{\mathrm{real}})
\approx_c
(\mathsf{view}_{P}^{\mathrm{sim}},\mathsf{out}_{\bar P}^{\mathrm{ideal}}),
\]
where $\bar P$ denotes the honest party. Modular composition extends the result to executions in which the resulting ciphertext handle or additive shares are consumed by secure subsequent algorithms.
\end{proof}

The proof also clarifies the implementation conditions that are essential for security: evaluation keys may be public, but decryption keys may not; server-to-client ciphertexts must be decrypted only after the prescribed fresh server mask has been applied; client-side correction masks must remain encrypted from the server; and masks must not be reused across unrelated revealed messages. Under these conditions, reducing ciphertext storage and replacing online ciphertext transfers by masked plaintext-sized messages does not introduce additional leakage beyond the stated output interfaces and public metadata.

\section{Fused Relinearization and Rotation}\label{Fused Relinearization and Rotation}

The previous two sections introduced preprocessing-based techniques that shift  ciphertext matrix multiplications in Transformer attention from the online phase to preprocessing, thereby reducing the online computation cost in two-party secure inference. These techniques can be applied to all Transformer layers in hybrid FHE-MPC systems and to the first layer in fully FHE-based systems. However, ciphertext matrix multiplications remain unavoidable in two important cases: (i) the preprocessing phase itself, where these computations are performed in advance, and (ii) the subsequent layers of fully FHE-based systems, where the proposed preprocessing techniques are not directly applicable. This motivates a deeper optimization of the underlying ciphertext matrix multiplication primitive itself.

In this section, we present a fused key-switching optimization that reduces the cost of ciphertext multiplication followed by rotation, a common computation pattern in secure Transformer inference systems such as BOLT~\cite{bolt}, THOR~\cite{Thor}, BLB~\cite{blb}, and ARION~\cite{ARION}. This optimization is independent of the proposed attention-computation algorithms and can be directly integrated into existing systems. When instantiated together with our preprocessing-based methods, it further accelerates ciphertext matrix multiplications in both preprocessing and online inference.

\subsection{Key-Switching Overhead in Multiplication-Rotation Pipelines}

In CKKS, ciphertext-ciphertext multiplication of
\[
\mathbf{ct}=(c_0,c_1)=Enc(m),
\qquad
\mathbf{ct}'=(c'_0,c'_1)=Enc(m')
\]
produces a degree-two ciphertext
\[
(d_0,d_1,d_2)
=
(c_0c'_0,\;c_0c'_1+c_1c'_0,\;c_1c'_1),
\]
which decrypts as
\[
d_0+d_1s+d_2s^2.
\]
A conventional implementation first relinearizes this ciphertext:
\[
d_0+d_1s+d_2s^2
\longrightarrow
\bar d_0+\bar d_1s \approx mm'.
\]
If a rotation by $t$ slots is then required, the implementation applies the automorphism $\sigma_t$ (rotation  left or right):
\[
(\sigma_t(\bar d_0),\sigma_t(\bar d_1)).
\]
This ciphertext is encrypted under the rotated secret key
\[
\tau=\sigma_t(s).
\]
Therefore, another key-switching operation is required:
\[
\sigma_t(\bar d_0)+\sigma_t(\bar d_1)\tau
\longrightarrow
c''_0+c''_1 s \approx \sigma_t(mm').
\]
Thus, multiplication followed by rotation normally invokes two key-switching procedures: one for relinearization and one for rotation.

Each key-switching operation consists of four main steps: ModUp, multiplication with the evaluation key (MulEvk), ModDown, and addition in RNS-CKKS~\cite{rns-ckks}. Since polynomial multiplication is most efficiently performed in the NTT domain, ciphertext multiplications are typically maintained in the NTT representation throughout the computation. Consequently, both ModUp and ModDown involve  NTT and inverse NTT (INTT) transformations. Furthermore, these two steps require basis conversion between different RNS bases, making them among the most expensive components of the key-switching procedure \cite{bossuat2021efficient}.

The key idea of our fused method is to postpone relinearization until after the automorphism operation. Instead of performing separate key-switching procedures, we directly transform the ciphertext encrypted under the secret-key $\tau$ and $\tau^2$ back to the original secret key $s$ through a single fused key-switching pipeline.

\subsection{Formulation}

After multiplication and rescaling, let
\[
(\tilde d_0,\tilde d_1,\tilde d_2)
\]
be the degree-two ciphertext at the target scale and modulus level. Applying the automorphism $\sigma_t$ gives
\[
(\hat d_0,\hat d_1,\hat d_2)
=
(\sigma_t(\tilde d_0),\sigma_t(\tilde d_1),\sigma_t(\tilde d_2)).
\]
This ciphertext decrypts under
\[
(1,\tau,\tau^2),
\qquad
\tau=\sigma_t(s),
\]
because
\[
\sigma_t(\tilde d_0+\tilde d_1s+\tilde d_2s^2)
=
\hat d_0+\hat d_1\tau+\hat d_2\tau^2.
\]
The goal is to produce a standard CKKS ciphertext under the original secret key $s$:
\[
\mathbf{ct}_{\mathrm{out}}=(c''_0,c''_1),
\qquad
c''_0+c''_1s
\approx
\hat d_0+\hat d_1\tau+\hat d_2\tau^2.
\]
This can be achieved if we have evaluation keys for both mappings:
\[
\tau\rightarrow s,
\qquad
\tau^2\rightarrow s.
\]
Instead of applying two separate key switches, we define a fused evaluation key
\[
\mathsf{evk}_{\mathrm{fused}}^{(i)}
=
\{
\mathsf{evk}_{\tau\rightarrow s}^{(i)},
\mathsf{evk}_{\tau^2\rightarrow s}^{(i)}
\} \quad i=0,1, \cdots,k+l-1.
\]
The RNS (Residue Number System) decomposition \cite{rns-ckks} of $\hat d_1$ and $\hat d_2$ is given by
\[
RNS_{D}(\hat d_1)=\{ \hat d_1^{(0)}, \hat d_1^{(1)}, \cdots , \hat d_1^{(k+l-1)} \}_D,
\]
\[
RNS_{D}(\hat d_2)=\{ \hat d_2^{(0)}, \hat d_2^{(1)}, \cdots , \hat d_2^{(k+l-1)} \}_D,
\]
where $D$ is a basis consisting of $k+l$ primes. Then, the ciphertext-evk (evaluation key) multiplication is given by:
\[
\check{ct}^{(i)} 
=
\hat d_1^{(i)}\mathsf{evk}_{\tau\rightarrow s}^{(i)}
+
\hat d_2^{(i)}\mathsf{evk}_{\tau^2\rightarrow s}^{(i)}.
\]
After one ModDown stage, ciphertext $\check{ {\mathbf{ct}}}=(\check{ct}_0,\check{ct}_1)$ satisfying
\[
\check{ct}_0+\check{ct}_1 s
\approx
\hat d_1\tau+\hat d_2\tau^2.
\]
The final ciphertext is
\[
\mathbf{ct}_{\mathrm{out}}
=
(\hat d_0+\check{ct}_0,\;\check{ct}_1),
\]
which decrypts to
\[
\hat d_0+\check{ct}_0+\check{ct}_1 s
\approx
\hat d_0+\hat d_1\tau+\hat d_2\tau^2.
\]

Therefore, the fused procedure jointly performs ciphertext multiplication and rotation while requiring only a single ModDown stage. The complete procedure is summarized in Algorithm~\ref{alg:fused-relin-rot}. It is worth noting that, in RNS-CKKS, ciphertexts are represented and manipulated in their RNS-decomposed form throughout the computation.

\begin{algorithm}[t]
\caption{ Fused Relinearization and Rotation}
\label{alg:fused-relin-rot}
\begin{algorithmic}[1]
\Require ciphertexts $RNS_C(\mathbf{ct})$ and $RNS_C(\mathbf{ct}')$, $C$ is a basis consisting of $l+1$ primes.
\Require Rotation step $t$ and fused evaluation key
\[
\mathsf{evk}_{\mathrm{fused}}^{(i)}
=
\left(
\mathsf{evk}_{\tau\rightarrow s}^{(i)},
\mathsf{evk}_{\tau^2\rightarrow s}^{(i)}
\right),
\quad
\tau=\sigma_t(s)
\]
\Ensure Ciphertext encrypting $\mathrm{Rot}_t(\mathbf{ct}\cdot\mathbf{ct}')$
\State Compute the degree-two product:
\[
RNS_C(d_0,d_1,d_2)
\leftarrow
RNS_C(c_0c'_0,\;c_0c'_1+c_1c'_0,\;c_1c'_1).
\]

\State Rescale without NTT:
\[
RNS_C(\tilde d_0,\tilde d_1,\tilde d_2)
\leftarrow
\mathsf{Rescale}(d_0,d_1,d_2)
\] One prime is removed from C.

\State Apply automorphism:
\[
RNS_C(\sigma_t(\tilde d_0),\sigma_t(\tilde d_1),\sigma_t(\tilde d_2))
\leftarrow
RNS_C(\tilde d_0,\tilde d_1,\tilde d_2).
\]

\State ModUp:
\[
RNS_D(\hat d_1)
\leftarrow
RNS_C(\sigma_t(\tilde d_1)).
\]
\[
RNS_D(\hat d_2)
\leftarrow
RNS_C(\sigma_t(\tilde d_2)).
\]

\State Initialize empty ciphertext $RNS_D(\check{ {\mathbf{ct}}})$.
\For{$i=0,\ldots,k+l-1$}
\State MulEvk:
\[
\check{ct}^{(i)} 
=
\hat d_1^{(i)}\mathsf{evk}_{\tau\rightarrow s}^{(i)}
+
\hat d_2^{(i)}\mathsf{evk}_{\tau^2\rightarrow s}^{(i)}.
\]
\EndFor
\State Perform a single ModDown:
\[
RNS_C(\check{ {\mathbf{ct}}})\leftarrow RNS_D(\check{ {\mathbf{ct}}}).
\]
\State Output
\[
RNS_C(\mathbf{ct}_{\mathrm{out}})=RNS_C(\sigma_t(\tilde d_0),0) + RNS_C(\check{ {\mathbf{ct}}}).
\]
\State \Return $RNS_C(\mathbf{ct}_{\mathrm{out}})$
\end{algorithmic}
\end{algorithm}

\subsection{Correctness}

We denote by $\mathsf{KS}_{a\rightarrow s}(x)$ the key-switching operation that transforms the ciphertext component $x$, associated with the secret-key $a$, into an equivalent component under the standard secret key $s$. In conventional CKKS, the multiplication-then-rotation operation is equivalent to
\[
\mathsf{KS}_{\tau\rightarrow s}
\left(
\sigma_t
\left(
\mathsf{KS}_{s^2\rightarrow s}
(d_2)
+
d_0+d_1s
\right)
\right).
\]
By contrast, the fused procedure computes
\[
\hat d_0+
\mathsf{KS}_{\tau\rightarrow s}(\hat d_1)
+
\mathsf{KS}_{\tau^2\rightarrow s}(\hat d_2).
\]

 Since key switching is linear in the decomposed input component, we have
\[
\mathsf{KS}_{\tau\rightarrow s}(\hat d_1)
+
\mathsf{KS}_{\tau^2\rightarrow s}(\hat d_2)
=
\mathsf{KS}_{\mathrm{fused}}
(\hat d_1,\hat d_2).
\]
Thus, the fused procedure preserves the correctness of standard CKKS key switching.

\subsection{Cost Advantage}

 A conventional multiplication followed by rotation requires approximately
\[
2\mathsf{ModUp}+2\mathsf{MulEvk}+2\mathsf{ModDown},
\]
ignoring low-cost automorphism and addition operations. The fused procedure requires
\[
2\mathsf{ModUp}+2\mathsf{MulEvk}+\mathsf{ModDown}.
\]
Therefore, one ModDown stage is eliminated.

\section{Evaluation}\label{evaluation}

To demonstrate the effectiveness of the general linear-layer preprocessing framework, we compare existing secure Transformer inference schemes with their counterparts after incorporating our algorithms, using attention as the common case study. Specifically, we consider two representative schemes: Arion, an FHE-based secure inference scheme, and BLB, a hybrid FHE-MPC secure inference scheme. For each scheme, we evaluate three variants. The original schemes are denoted by Arion and BLB. Arion(1) and BLB(1) denote the schemes after integrating the preprocessing-based attention computation in Algorithm~\ref{alg:secure-attention}. Arion(2) denotes the storage-reduced FHE-output algorithm in Algorithm~\ref{tradeoffs2s3FHE}, while BLB(2) denotes the hybrid FHE-output algorithms in Algorithms~\ref{tradeoffs2s3} and~\ref{tradeoffS5}.

Unless otherwise specified, all evaluations are conducted on the attention computation of the {BERT-base} model. Following the settings adopted in both Arion and BLB, the inference input consists of a batch of token embeddings containing $m=128$ tokens. Since all attention heads in a multi-head attention layer share the same computation pattern, the statistics reported in Table~\ref{rotationcompare} correspond to the computation of a {single attention head}.

Table~\ref{rotationcompare} summarizes the impact of the proposed algorithms on key-switching operations, preprocessing storage overhead, and online communication costs for the three major computations in the attention mechanism. For Arion, we apply Algorithm~\ref{alg:secure-attention} only to the first Transformer layer during inference and do not use the baby-step giant-step method adopted in the original Arion design. For BLB, before integrating our algorithms, we adjust the position of the MPC-to-CKKS conversion algorithm so that the data representation matches the inputs required by our secure attention computation.

\begin{table*}[ht]
\caption{Effects of the proposed techniques on key-switching operations, preprocessing storage, and online communication in BLB and Arion. }
\label{rotationcompare}
\centering
\small
\renewcommand{\arraystretch}{1.5}
\begin{tabularx}{\linewidth}{>{\centering\arraybackslash}m{0.1667\textwidth} 
>{\centering\arraybackslash}m{0.1667\textwidth} 
>{\centering\arraybackslash}m{0.0667\textwidth} 
>{\centering\arraybackslash}m{0.0667\textwidth} 
>{\centering\arraybackslash}m{0.1667\textwidth} 
>{\centering\arraybackslash}m{0.1667\textwidth}}
\toprule
Operation & Scheme &Slot & Key Switch & \shortstack{Ciphertext Storage \\ in Preprocess Stage}  & Data Traffic(online) \\
\midrule
\multirow{5}{=}{\centering Q/K/V Projection \\ $\{XW_*\}^{(h)}_{h \in [H]}$ \\ $X \in \mathbb{R}^{m \times n}$ \\ $W_* \in \mathbb{R}^{n \times \frac{n}{H}}$ \\ $* = Q,K,V$   }
& Arion & $2^{15}$ &0 & 0& $\mathbf{X}-\mathbf{U}$ \\
& Arion(1) & $2^{15}$ &0 & 0&0  \\
& Arion(2) & $2^{15}$ &0 & 0&0  \\
& BLB & $2^{14}$ &192 & 0&  $[[ \left \langle \mathbf{X} -\mathbf{U} \right \rangle _{c}^{q}]]$  \\
& BLB(1) & $2^{14}$ &0 & 0&0  \\
& BLB(2) & $2^{14}$ &0 & 0&0  \\

\midrule
\multirow{5}{=}{\centering Attention Scores \\ $\{Q^{(h)}(K^{(h)})^\intercal\}_{h \in [H]}$ \\ $Q^{(h)} \in \mathbb{R}^{m \times \frac{n}{H}}$ \\ $K^{(h)} \in \mathbb{R}^{m \times \frac{n}{H}}$ }
& Arion & $2^{15}$ &1536 & 0 & 0 \\ 
& Arion(1) & $2^{15}$ &0 & 8384 & $\mathbf{X}-\mathbf{U}$  \\
& Arion(2) & $2^{15}$ &0 & 256 & $\mathbf{X}-\mathbf{U}$, $\mathbf{B_1}+\mathbf{B_2}$  \\
& BLB & $2^{14}$ &346 & 0&  0  \\
& BLB(1) & $2^{14}$ & 256 & 50 & $ \left \langle \mathbf{X} \right \rangle _{c} -\mathbf{U}$  \\
& BLB(2) & $2^{14}$ &0 & 4& $\left \langle \mathbf{X} \right \rangle _{c}-\mathbf{U}$, $\mathbf{E}+\mathbf{F}$   \\
\midrule
\multirow{5}{=}{\centering  Value Aggregation \\ $\sigma = softmax$ \\ $\{\sigma(Q^{(h)}(K^{(h)})^\intercal)V^{(h)}\}_{h \in [H]}$ \\ $\sigma(Q^{(h)}(K^{(h)})^\intercal) \in \mathbb{R}^{m \times m}$ \\ $V^{(h)} \in \mathbb{R}^{m \times \frac{n}{H}}$}
& Arion & $2^{15}$ &2432 & 0 & 0 \\ 
& Arion(1) & $2^{15}$ &64 & 8192 & 0 \\
& BLB & $2^{14}$ &128 & 0&  $[[ \left \langle \mathbf{X'} -\mathbf{U'} \right \rangle _{c}]]$  \\
& BLB(1) & $2^{14}$ & 0 & 257 & $ \left \langle \mathbf{X'} \right \rangle _{c}  -\mathbf{U'}$  \\
& BLB(2) & $2^{14}$ &0 & 2& $\left \langle \mathbf{X'}  \right \rangle _{c}-\mathbf{U'}$, $\mathbf{J}$, $\mathbf{P}$   \\
\bottomrule
\end{tabularx}
\end{table*}

For the Q/K/V projection stage,  the online key-switching operations required by the original BLB scheme, are completely eliminated. The projection matrices are instead generated through the masked preprocessing procedure described in Algorithm~\ref{alg:secure-attention}. Since Arion already performs this computation without online key switching, the algorithm does not introduce additional overhead in this stage.

The largest improvements are observed in the computation of attention scores. For the original Arion scheme, evaluating
\(
Q^{(h)}(K^{(h)})^\top
\)
requires 1536 key-switching operations. After applying Algorithm~\ref{alg:secure-attention}, all of these operations are removed by precomputing ciphertext transformation sets during preprocessing. However, this optimization introduces a storage overhead of 8384 ciphertexts. Algorithm~\ref{tradeoffs2s3FHE} further reduces this storage requirement to only 256 ciphertexts by replacing transformed ciphertexts with the online transmission of the auxiliary values $\mathbf{B_1}+\mathbf{B_2}$. A similar trend can be observed for BLB. The original scheme requires 346 key-switching operations, which are reduced to 256 by Algorithm~\ref{alg:secure-attention} and completely eliminated by Algorithm~\ref{tradeoffs2s3}. At the same time, the preprocessing storage overhead decreases from 50 ciphertexts to only 4 ciphertexts, with the additional communication of $\mathbf{E}+\mathbf{F}$ during online inference.

The value aggregation stage exhibits similar behavior. For Arion, the original implementation requires 2432 key-switching operations, while Algorithm~\ref{alg:secure-attention} reduces this number to 64. These remaining operations are associated with ciphertext transformations required for the value matrix. In the hybrid BLB scheme, the original 128 key-switching operations are completely removed. Furthermore, Algorithm~\ref{tradeoffS5} reduces the preprocessing storage requirement from 257 ciphertexts to only 2 ciphertexts by replacing the stored transformed ciphertexts with the online transmission of $\mathbf{J}$ and $\mathbf{P}$.

The communication entries in Table~\ref{rotationcompare} should be interpreted together with the representation being transferred. The original hybrid pipeline transmits FHE ciphertexts at encrypted linear-layer boundaries, whereas the covered online linear-layer procedures of the proposed hybrid algorithms transmit no FHE ciphertext in either direction. Their online messages are masked shares or masked correction matrices with plaintext-sized element counts. This difference is not visible from operation counts alone, but it is important in bandwidth-constrained deployment because FHE ciphertexts are substantially larger. For BLB(2), retaining an FHE output is useful because BLB can continue homomorphic computation within subsequent nonlinear functions, allowing the integrated pipeline to avoid both interactive ring-to-field and field-to-ring conversions. If the following nonlinear stage is conventional MPC, Algorithms~\ref{tradeoffMPCQK} and~\ref{tradeoffMPCMHA} instead return additive shares directly. They also further reduce online communication: the former retains $\mathbf{E}+\mathbf{F}$ locally rather than transmitting it, and the latter sends $\mathbf{J}$ without the additional correction $\mathbf{P}$ required by its FHE-output counterpart.

Overall, the results demonstrate four complementary advantages. First, Algorithm~\ref{alg:secure-attention} substantially reduces online key switching, and the storage-reduced variants eliminate it for the covered linear computations. Second, replacing ciphertext transformation sets with masked corrections greatly reduces preprocessing storage. Third, the hybrid online boundary uses plaintext-sized masked messages without FHE ciphertext transmission in either direction, avoids both interactive ring-to-field and field-to-ring conversions at the corresponding interface, and offers still lower online communication when secret-shared outputs are selected. Fourth, these changes do not prescribe a new packing method: the same algorithm structure can be instantiated with different compatible encodings. The evaluation therefore supports the central claim that preprocessing is a general linear-layer interface technique, while the attention case study quantifies its computational and storage effects in representative systems.

\section{Conclusion}\label{conclusion}

This paper presented a general linear-layer preprocessing framework for two-party privacy-preserving Transformer inference. The construction moves input-independent encrypted computation offline and retains only masked matrix operations online. Since the algorithms are specified through matrix-level relations rather than slot assignments or packing-specific rotation schedules, they support compatible CKKS, BFV, and BGV instantiations without requiring a new encoding or packing strategy.

The attention constructions provide both FHE and additive-secret-shared outputs, allowing existing homomorphic and MPC nonlinear protocols to be retained. Their storage-reduced variants replace cached ciphertext transformation sets with masked corrections, eliminate online key switching for the covered linear computation, and avoid online FHE ciphertext transmission in hybrid inference. At matching interfaces, they also avoid interactive ring-to-field and field-to-ring conversions. For an existing end-to-end system, integration primarily requires replacing the covered linear-layer preprocessing and online message flow; the packing strategy, model, and downstream nonlinear protocols can remain unchanged. The resulting improvements therefore require localized modifications rather than a system-wide redesign. 

A fused relinearization-and-rotation procedure further reduces the cost of multiplication followed by rotation in RNS-CKKS. Correctness and security against static semi-honest adversaries were established under the stated RLWE and masking assumptions. Our case studies show that the proposed methods can be applied to existing end-to-end inference systems with only minor modifications, reducing key switching, preprocessing storage, and online communication.

\bibliographystyle{unsrt}
\bibliography{cas-refs}
\vfill
\end{document}